\documentclass{article}%
\usepackage{amsmath}
\usepackage{amsfonts}
\usepackage{amssymb}
\usepackage{graphicx}%
\usepackage{caption}
\providecommand{\U}[1]{\protect\rule{.1in}{.1in}}
\newtheorem{theorem}{Theorem}[section]

\newtheorem{definition}{Definition}[section]
\newtheorem{example}{Example}[section]

\newtheorem{lemma}[theorem]{Lemma}
\newtheorem{proposition}[theorem]{Proposition}
\newtheorem{remark}{Remark}[section]

\begin{document}

\title{Optimal ratcheting of dividends in insurance}
\author{Hansj\"{o}rg Albrecher\thanks{Department of Actuarial Science, Faculty of
Business and Economics, University of Lausanne, CH-1015 Lausanne and Swiss
Finance Institute. Supported by the Swiss National Science Foundation Project
$200021\_168993$.}, Pablo Azcue\thanks{Departamento de Matematicas,
Universidad Torcuato Di Tella. Av. Figueroa Alcorta 7350 (C1428BIJ) Ciudad de
Buenos Aires, Argentina.} and Nora Muler$^{\dag}$}
\date{}
\maketitle

\bigskip

\abstract{\begin{quote}We address a long-standing open problem in risk theory, namely the optimal strategy to pay out dividends from an insurance surplus process, if the dividend rate can never be decreased. The optimality criterion here is to maximize the expected value of the aggregate discounted dividend payments up to the time of ruin. In the framework of the classical Cram\'{e}r-Lundberg risk model, we solve the corresponding two-dimensional optimal control problem and show that the value function is the unique viscosity solution of the corresponding Hamilton-Jacobi-Bellman equation. We also show that the value function can be approximated arbitrarily closely by ratcheting strategies with only a finite number of possible dividend rates and identify the free boundary and the optimal strategies in several concrete examples. These implementations illustrate that the restriction of ratcheting does not lead to a large efficiency loss when compared to the classical un-constrained optimal dividend strategy.\end{quote}}

\section{Introduction}

How to optimally pay out dividends from an insurance surplus process is a
classical research question starting with the papers of de Finetti \cite{defin} and
Gerber \cite{Ger69}. When the criterion is to maximize the expected aggregate
discounted dividend payments up to the time of ruin, the challenge is to find
the right compromise between paying early in view of the discounting and
paying late in order not to have ruin too early and profit from the typically
positive safety loading for a longer time. The problem turns out to be very
challenging from a mathematical point of view, and many variants have been
studied over the last decades, using various different techniques, see
e.g.\ Schmidli \cite{Schmidli book 2008} and Albrecher \& Thonhauser \cite{AT}
for an overview. In recent years, the problem became well understood within
the framework of modern stochastic control theory and the concept of
viscosity solutions for corresponding Hamilton-Jacobi-Bellman equations,
cf.\ Azcue \& Muler \cite{AM Libro}.\newline

In terms of the practical insight from the resulting optimal payout
strategies, one aspect often raised critically in discussions by practitioners
was the following: dividend strategies implemented in practice often are
designed in a way as to not decrease over time, since a decrease would send
unfavorable signals to the market. Such a monotonicity of dividend rates over
time (also referred to as \textit{ratcheting}) is, however, not automatically
present in the optimal strategies without this ratcheting constraint, as the
optimal strategies are of band type (and often of simpler threshold form: pay
no dividends below a certain threshold, and at maximal rate above the
threshold). Hence it is an interesting question to (a) look for the optimal
strategies when such a ratcheting constraint is imposed and (b) see whether
this additional constraint comes at the cost of losing a lot of efficiency
when compared to the un-constrained value function.\footnote{The question was
for instance posed in an academic environment by Elias Shiu at the First Int. Workshop on Gerber-Shiu functions in
Montreal back in 2006, see also Avanzi et al.\ \cite{aw} for a more recent
motivation.} A first step towards answering these questions was recently
obtained in \cite{ABB}, where explicit calculations were performed for a
restricted form of a ratcheting strategy, namely that once during the lifetime
of the process the dividend rate can be increased. It was then studied, both
in the Cram\'{e}r-Lundberg model and its diffusion approximation, to what
extent and at which surplus level such an increase should optimally be
implemented, leading to some surprising relations of the optimal ratcheting
level with the threshold level of unconstrained dividend strategies. However,
finding the optimal solution to the general ratcheting problem for a continuum
of available ratcheting levels of the dividend rate was still open. From a
technical point of view, it becomes clear that in a stochastic control
formulation one is faced with a (Markovian) two-dimensional problem, keeping
track of both the current surplus level and the currently implemented dividend
rate. The analysis of two-dimensional control problems in risk theory can be
quite intricate, see e.g.\ Albrecher et al.\ \cite{AAM} and Gu et
al.\ \cite{Gu} in the context of other dividend problem formulations. \newline

In this paper, we solve the two-dimensional ratcheting problem and establish the value
function as the unique viscosity solution of the respective
Hamilton-Jacobi-Bellman equation. It will turn out that allowing the maximal
dividend rate to exceed the rate of incoming premiums leads to some additional
analytical challenges, but one can derive the respective results in that case
as well. Note that the concept of ratcheting has been studied in the context
of lifetime consumption in the corporate finance community, see Dybvig
\cite{dybvig} and the very recent nice extension of Angoshtari et
al.\ \cite{bayr}. In those papers the focus is on a geometric Brownian motion
as an underlying and a logarithmic or power utility function applied to the
consumption rate, which together with interest rate considerations renders
this model setup within the framework of Merton-type consumption problems.
Despite some apparent analogies, the present risk theory setup does not fall
within the class of models studied there and the techniques used for its study
are quite different. \newline

The rest of the paper is structured as follows. Section \ref{sec2} describes
the model setup in more detail and some basic results are derived in Section
\ref{sec3}. Some of the respective proofs are, however, quite technical and hence
delegated to an appendix. In Section \ref{sec4} it is then proved that the
value function of the general ratcheting problem is the unique viscosity
solution of the respective Hamilton-Jacobi-Bellman equation. Section
\ref{sec5} studies properties of ratcheting strategies when only finitely many
different dividend rates are possible, and in Section \ref{sec6} it is shown
that these strategies converge uniformly to the general value function, when
the number of possible dividend rates tends to infinity. Section
\ref{Seccion Ejemplos} identifies the resulting optimal ratcheting strategies,
the free boundaries and the corresponding value functions for a number of
concrete examples with exponentially and Gamma distributed claims. We also
compare the optimal solutions to their counterparts in the un-constrained case
(without ratcheting) and in the case when only one switch of dividend rate is
allowed, as studied in \cite{ABB}. It turns out that the efficiency loss due
to ratcheting is remarkably small, and that a one-switch strategy already performs
very similarly to the optimal general ratcheting solution. Finally, Section
\ref{seccon} concludes.

\section{Model}

\label{sec2}

Consider the free surplus $X_{t}$ of an insurance portfolio according to the
Cram\'{e}r-Lundberg model
\begin{equation}
X_{t}=x+pt-%
{\textstyle\sum\nolimits_{i=1}^{N_{t}}}
U_{i}, \label{freeSurplus}%
\end{equation}
where $x$ is the initial surplus, $p$ is the premium rate and $U_{i}\ $is the
size of the $i$-th claim. All claims are assumed to be i.i.d.\ random
variables with continuous distribution function $F$. $N_{t}\ $is the number of claims
up to time $t$ and assumed to follow a Poisson process with intensity $\beta$.
Let us denote by $\tau_{i}$ the arrival time of claim $i$. The process $N_{t}%
$\ and the random variables $U_{i}$ are independent of each other, and we have
the safety loading condition $p>\beta\mathbb{E}(U_{i})$. Let $\Omega$ be the
set of paths with left and right limits and let $(\Omega,\mathcal{F},\left(
\mathcal{F}_{t}\right)  _{t\geq0},\mathcal{P})$ be the complete probability
space generated by the process $X_{t}$.

The company uses part of the surplus to pay dividends to the shareholders with
a finite rate\ less than or equal to a fixed rate $\overline{c}>0$. Let us
denote by $C_{t}$ the rate at which the company pays dividends at time $t$.
Given an initial surplus $X_{0}=x$ and a minimum dividend rate $c$ at the
beginning, a dividend ratcheting strategy $C=\left(  C_{t}\right)  _{t\geq0}$
is admissible if it is c\`{a}dl\`{a}g, adapted with respect to the filtration
$\left(  \mathcal{F}_{t}\right)  _{t\geq0}$, non-decreasing and if it
satisfies $c\leq C_{t}\leq\overline{c}$ for all $t$. Moreover, the controlled
surplus process can be written as%
\[
X_{t}^{C}=X_{t}-\int_{0}^{t}C_{s}ds.
\]
Let us define $\Pi_{x,c,\overline{c}}$ as the set of all the admissible
dividend ratcheting strategies. Given $x\geq0$, $c\in\left[  0,\overline
{c}\right]  $ and an admissible dividend ratcheting strategy $C\in
\Pi_{x,c,\overline{c}}$, the value function of this strategy is given by%
\[
J(x;C)=\mathbb{E}\left[  \int_{0}^{\tau}e^{-qs}C_{s}ds\right]  ,
\]
where $\tau=\inf\left\{  t\geq0:X_{t}^{C}<0\right\}  $ is the ruin time. Hence,
for any initial surplus $x\geq0$ and initial dividend rate $c\in\left[
0,\overline{c}\right]  $, our aim is to maximize%

\begin{equation}
V(x,c)=\sup_{C\in\Pi_{x,c,\overline{c}}}J(x;C). \label{optimal value function}%
\end{equation}

\begin{remark}
\normalfont
\label{Remark Pagar p}In the case that $c\leq p,$ any ratcheting strategy $C$
in $\Pi_{0,c,\overline{c}}$ with $C_{0}>p$ can not be optimal because the
corresponding ruin time is $0.$ Also, in the case that $C_{t-}\leq p$ and
$X_{t}^{C}=0$ for some $t>0$, any dividend ratcheting strategy with $C_{s}>p$
for $s>t$ can not be optimal because it implies immediate ruin as well. So,
without loss of generality, we only consider admissible strategies that
satisfy the following property: if $C_{t^{-}}=p$ with $X_{t}^{C}=0$ for $t>0$,
then $C_{s}=p$ for $s\geq t$ until ruin time. Also, the only admissible
strategy in $\Pi_{0,p,\overline{c}}$ that we consider is to pay dividends at
rate $p$ up to the arrival of the first claim, which is the ruin time.
\end{remark}

\begin{remark}
\normalfont
\label{Optima sin ratcheting} The dividend optimization problem without the
ratcheting constraint, that is where the dividend strategy $C=\left(
C_{t}\right)  _{t\geq0}$ is not necessarily non-decreasing, was studied
intensively in the literature (see e.g.\ Gerber and Shiu \cite{Gerber 2006}, Schmidli
\cite[Sec.2.4]{Schmidli book 2008} and Azcue and Muler \cite{AM Bounded
2012}). Unlike the ratcheting optimization problem, this non-ratcheting
problem is one dimensional. If $V^{NR}(x)$ denotes the optimal value function
of this non-ratcheting problem, then clearly $V(x,c)\leq V^{NR}(x)$ for all
$x\geq0$ and $c\in\lbrack0,\overline{c}]$. It is known that $V^{NR}$ is
non-decreasing with $\lim_{x\rightarrow\infty}V^{NR}(x)=\overline{c}/q$.
Moreover, in \cite{AM Bounded 2012}, it was proved that there exists an
optimal strategy and it has a band structure. It is characterized by three
sets which partition the state space of the surplus process $[0,\infty)$. Each
set is associated with a certain dividend payment action: define
$\mathcal{O}^{NR}$ as the set of values where no dividends are paid,
$\mathcal{B}^{NR}$ as the set of values where dividends are paid at the
maximum possible rate $\overline{c}$ and $\mathcal{A}^{NR}$ as the set of
values where dividends are paid at rate $p$. The topological properties of
these sets depend on whether the premium rate $p$ is larger than the
dividend-rate ceiling $\overline{c}$; for example $\mathcal{A}^{NR}$ is empty
if $l_{0}<p$, and $\mathcal{B}^{NR}$ is empty in the case
$\overline{c}=p$. The band strategies are \textit{stationary} in the sense
that they only depend on the current surplus. The simplest band strategies are
the so-called \textit{threshold} strategies, according to which dividends are paid
at the maximal admissible rate $\overline{c}$ as soon as the surplus exceeds a
certain threshold level $x_{NR}\geq0$ and no dividends are paid when the
surplus is less than $x_{NR}$. More precisely, the threshold strategy is
characterized by the sets $\mathcal{O}^{NR}=[0,x_{NR})$ and $\mathcal{B}%
^{NR}=[x_{NR},\infty)$ in the case $\overline{c}<p$, by the sets
$\mathcal{O}^{NR}=[0,x_{NR})$ and $\mathcal{A}^{NR}=[x_{NR},\infty)$ in the
case $\overline{c}=p$, and by the sets $\mathcal{O}^{NR}=[0,x_{NR})$,
$\mathcal{A}^{NR}=\{x_{NR}\}$ and $\mathcal{B}^{NR}=(x_{NR},\infty)$ in the
case $\overline{c}>p$.
\end{remark}

\section{Basic Results}

\label{sec3}

Let us first derive some basic properties of the optimal value function
(\ref{optimal value function}). In the case $\overline{c}\leq p$ we will show
that the optimal value function $V$ is globally Lipschitz. In contrast, for
$\overline{c}>p$ there are some issues with the regularity and the proofs are
more involved. It is clear that in the case of $\overline{c}>p,$ the optimal
value function $V$ is not continuous at the point $(0,p)$. Indeed, by Remark
\ref{Remark Pagar p}, we have that
\[
V(0,p)=\mathbb{E}\left[  \int_{0}^{\tau_{1}}e^{-qs}p\,ds\right]  =\frac
{p}{q+\beta};
\]
but $V(0,c)=0$ for all $c\in(p,\overline{c}]$ because all the admissible
strategies lead to immediate ruin. As a consequence,
\[
\lim_{c\to p^{+}}V(0,c)=0<V(0,p)=p/(q+\beta).
\]

In the case $\overline{c}>p,$ we prove the following results depending on the
value of $c$: (1) $V$ is Lipschitz in $\ [0,\infty)\times\lbrack0,p]$, (2) $V$
is continuous with respect to $c$ in $(0,\infty)\times\left\{  p\right\}  $.
(3) $V$ is locally Lipschitz in $[0,\infty)\times(p,\overline{c}]$ with a
Lipschitz bound that goes to infinity as $c\searrow p.$ In particular, we
conclude that $V$ is continuous at any point except $(0,p).$

Let us start with a straightforward result regarding the boundedness and
monotonicity of the optimal value function.

\begin{proposition}
\label{Monotone Optimal Value Function}The optimal value function $V(x,c)$ is
bounded by $\overline{c}/q$, non-decreasing in $x$ and
non-increasing in $c.$
\end{proposition}

\emph{Proof.} Since the discounted value of paying the maximum rate
$\overline{c}$ up to infinity is $\overline{c}/q,$ we conclude the boundedness result.

On the one hand $V(x,c)$ is non-increasing in $c$ because given $c_{1}<c_{2}$
we have $\Pi_{x,c_{2},\overline{c}}\subset\Pi_{x,c_{1},\overline{c}}$ for any
$x\geq0$. On the other hand, given $x_{1}<x_{2}$ and an admissible ratcheting
strategy $C^{1}\in\Pi_{x_{1},c,\overline{c}}$ for any $c\in\left[
0,\overline{c}\right]  $, let us define $C^{2}\in\Pi_{x_{2},c,\overline{c}}$
as $C_{t}^{2}=C_{t}^{1}$ until the ruin time of the controlled process
$X_{t}^{C^{1}}$ with $X_{0}^{C^{1}}=x_{1}$, and pay the maximum rate
$\overline{c}$ afterwards. Thus, $J(x;C_{1})\leq J(x;C_{2})$ and we have the
result.\hfill $\blacksquare$\newline

Note that the previous proposition implies
\[
0\leq V(x_{2},c_{1})-V(x_{1},c_{2})
\]
for all $0\leq x_{1}\leq x_{2}$ and $c_{1}\leq c_{2}$.

In order to obtain the Lipschitz results we add the following assumption for
technical reasons:

\begin{quote}A1. The claim size distribution $F$ is (globally) Lipschitz, that is if $x<y,
$ $0\leq F(y)-F(x)\leq K(y-x)$ for some $K>0$.\end{quote}

The following proposition establishes that $V$ is Lipschitz in the case
$\overline{c}\leq p$ and also in the case $\overline{c}>p$ for $(x,c)\in
\lbrack0,\infty)\times\lbrack0,p].$

\begin{proposition}
\label{Proposition Global Lipschitz zone}There exists a constant $K_{1}>0$
such that
\[
0\leq V(x_{2},c_{1})-V(x_{1},c_{2})\leq K_{1}\left[  \left(  x_{2}%
-x_{1}\right)  +\left(  c_{2}-c_{1}\right)  \right]
\]

for all $0\leq x_{1}\leq x_{2}$ and $c_{1}\leq c_{2}\leq\min\left\{
\overline{c},p\right\}  .$
\end{proposition}

The proof of this proposition is in the Appendix.

Lipschitz bounds for the case
$\overline{c}>p$ with $(x,c)\in\lbrack0,\infty)\times(p,\overline{c}]$ are as follows:

\begin{proposition}
\label{Proposition Lipschitz cmax mayor a p} Assume that $\overline{c}>p$,
then there exist constants $K_{2\text{ }}>0$ and $K_{3}>0$ such that%
\[
0\leq V(x_{2},c_{1})-V(x_{1},c_{2})\leq\left[  K_{2}+\frac{K_{3}}{c_{1}%
-p}\right]  \left(  x_{2}-x_{1}\right)  +\left[  K_{2}+\frac{K_{3}x_{2}%
}{\left(  c_{1}-p\right)  ^{2}}\right]  \left(  c_{2}-c_{1}\right)
\]
for all $0\leq x_{1}\leq x_{2}$ and $p<c_{1}\leq c_{2}\leq\overline{c}.$
\end{proposition}

The proof of this proposition is in the Appendix.

Note that in the case $\overline{c}>p$, from Proposition
\ref{Proposition Global Lipschitz zone} we have a global Lipschitz condition
in $[0,\infty)\times\lbrack0,p]$ for $V$ and Proposition \ref{Proposition Lipschitz cmax mayor a p} guarantees a local Lipschitz condition in $\ [0,\infty)\times(p,\overline{c}]$. The
next proposition deals with the continuity from above of $V$ in the set
$(0,\infty)\times\{p\}$.

\begin{proposition}
\label{Continuidad en p en la variable c} Assume $\overline{c}>p$, then we
have that $\lim_{c\rightarrow p^{+}}V(x,c)=V(x,p)$ for $x>0.$
\end{proposition}

The proof of this proposition is in the Appendix.

\section{Viscosity Solutions}

\label{sec4}

In this section we introduce the Hamilton-Jacobi-Bellman (HJB) equation of the
ratcheting problem and show that, in some sense, the optimal value function
$V$ defined in (\ref{optimal value function}) is the unique viscosity solution
of the HJB equation with boundary condition $\overline{c}/q$ when $x$ goes to
infinity. In the case that $\overline{c}\leq p,$ we will prove that the
optimal value function $V$ is the unique viscosity solution in $(0,\infty
)\times(0,\overline{c}]$ satisfying $\lim_{x\rightarrow\infty}V(x,c)=\overline
{c}/q$. For $\overline{c}>p$, the scenario is more complex: we first prove
that $V$ is the unique viscosity solution in $(0,\infty)\times(p,\overline
{c}]$ satisfying $\lim_{x\rightarrow\infty}V(x,c)=\overline{c}/q$ and
afterwards that $V$ is the unique viscosity solution in $(0,\infty
)\times\lbrack0,p]$ satisfying $V(x,p)=\lim_{c\rightarrow p^{+}}V(x,c)$ for
$x>0$ (here, we use the continuity result of Proposition
\ref{Continuidad en p en la variable c}).

Let us define the operator%
\begin{equation}
\mathcal{L}(u)(x,c)=c+(p-c)u_{x}(x,c)-(q+\beta)u(x,c)+\beta\int\nolimits_{0}%
^{x}u(x-\alpha,c)dF(\alpha). \label{Operador L}%
\end{equation}

The Hamilton-Jacobi-Bellman equation associated to
(\ref{optimal value function}) is given by%
\begin{equation}
\max\{\mathcal{L}(u)(x,c),u_{c}(x,c)\}=0\text{ for }x\geq0\ \text{and }0\leq
c\leq\overline{c}\text{.} \label{HJB equation}%
\end{equation}

\begin{definition}
\label{Viscosity} (a) A locally Lipschitz function $\overline{u}%
:[0,\infty)\times\lbrack c_{1},c_{2})\rightarrow\mathbb{R}$, where $0\leq
c_{1}<c_{2}\leq\overline{c}$, is a viscosity supersolution of
(\ref{HJB equation})\ at $(x,c)\in(0,\infty)\times\lbrack c_{1},c_{2})$\ if
any continuously differentiable function $\varphi:[0,\infty)\times\lbrack
c_{1},c_{2})\rightarrow\mathbb{R}\ $with $\varphi(x,c)=\overline{u}(x,c)$ such
that $\overline{u}-\varphi$\ reaches the minimum at $\left(  x,c\right)
$\ satisfies
\[
\max\left\{  \mathcal{L}(\varphi)(x,c),\varphi_{c}(x,y)\right\}  \leq0.\
\]
The function $\varphi$ is called a \textbf{test function for supersolution} at
$(x,c)$.

(b) A function $\underline{u}:$ $[0,\infty)\times\lbrack c_{1},c_{2}%
)\rightarrow\mathbb{R}\ $\ is a viscosity subsolution\ of (\ref{HJB equation}%
)\ at $(x,c)\in(0,\infty)\times\lbrack c_{1},c_{2})$\ if any continuously
differentiable function $\psi:[0,\infty)\times\lbrack c_{1},c_{2}%
)\rightarrow\mathbb{R}\ $with $\psi(x,c)=\underline{u}(x,c)$ such that
$\underline{u}-\psi$\ reaches the maximum at $\left(  x,c\right)  $ satisfies
\[
\max\left\{  \mathcal{L}(\psi)(x,c),\psi_{c}(x,c)\right\}  \geq0\text{.}%
\]
The function $\psi$ is called a \textbf{test function for subsolution} at
$(x,c)$.

(c) A function $u:[0,\infty)\times\lbrack c_{1},c_{2})$ which is both a
supersolution and subsolution at $(x,c)\in\lbrack0,\infty)\times\lbrack
c_{1},c_{2})$ is called a viscosity solution of (\ref{HJB equation})\ at
$(x,c)$.
\end{definition}

We first prove that $V$ is a viscosity solution of the HJB equation except at
the points of the set $(0,\infty)\times\{p\}$ where $V$ might not be locally
Lipschitz. Let us first state the dynamic programming principle. The proof is
similar to the one of Lemma 1.2 in \cite{AM Libro}.

\begin{lemma}
\label{DPP} Given any stopping time $\widetilde{\tau}$, we can write%
\[
V(x,c)=\sup\limits_{C\in\Pi_{x,c,\overline{c}}}\mathbb{E}\left[  \int
_{0}^{\tau\wedge\widetilde{\tau}}e^{-qs}C_{s}ds+e^{-q(\tau\wedge
\widetilde{\tau})}V(X_{\tau\wedge\widetilde{\tau}}^{C},C_{\widetilde{\tau}%
})\right]  \text{.}%
\]

\end{lemma}

\begin{proposition}
\label{Proposicion Viscosidad} (i) If $\overline{c}\leq p$, $V$ is a viscosity
solution of (\ref{HJB equation}) in $(0,\infty)\times\lbrack0,\overline{c})$.
(ii) If $\overline{c}>p$, $V$ is a viscosity solution of (\ref{HJB equation})
in $(0,\infty)\times\lbrack0,p)$ and also in $(0,\infty)\times\lbrack
c_{1},\overline{c})$ for any $c_{1}>p$.
\end{proposition}

\emph{Proof.} We prove here part (i). The proof of part (ii) is similar.

Let us first show that $V$ is a viscosity supersolution in $(0,\infty
)\times\lbrack0,\overline{c})$ for $\overline{c}\leq p$. By Proposition
\ref{Monotone Optimal Value Function}, $V_{c}\leq0$ in $(0,\infty
)\times\lbrack0,\overline{c})$ in the viscosity sense.

Consider now $(x,c)\in(0,\infty)\times\lbrack0,\overline{c})$ and the
admissible strategy $C\in\Pi_{x,c,\overline{c}}$ which pays dividends at
constant rate $c$ up to the ruin time $\tau$. Let us denote the corresponding
controlled surplus process as $X_{t}^{C}=X_{t}-ct$ and suppose that there
exists a test function $\varphi$ for supersolution (\ref{HJB equation}) at
$(x,c)$. This means that $\varphi$ is a continuously differentiable function
$\varphi:[0,\infty)\times\lbrack0,\overline{c}]\rightarrow\mathbb{R}\ $with
$\varphi(x,c)=V(x,c)$ and such that $V-\varphi$\ reaches the minimum at
$(x,c)$. \ We extend the definition of both $V$ and $\varphi$ as $\varphi=0$
for $x<0$. Using Lemma \ref{DPP} we get for $h>0$ ,%
\[%
\begin{array}
[c]{lll}%
\varphi(x,c) & = & V(x,c)\\
& \geq & \mathbb{E}\left[  \int\nolimits_{0}^{\tau_{1}\wedge h}e^{-q\,s}%
\,cds\right]  +\mathbb{E}\left[  e^{-q(\tau_{1}\wedge h)}\varphi(X_{\tau
_{1}\wedge h}^{C},c))\right]  .
\end{array}
\]
Hence,%
\[%
\begin{array}
[c]{lll}%
0 & \geq & \mathbb{E}\left[  \int\nolimits_{0}^{\tau_{1}\wedge h}%
e^{-q\,s}\,c\,ds\right]  +\mathbb{E}\left[  I_{\tau_{1}>h}e^{-q\,(\tau
_{1}\wedge h)}\varphi(x+(p-c)h,c)\right] \\
&  & +\mathbb{E}\left[  I_{\tau_{1}\leq h}e^{-q\,(\tau_{1}\wedge h)}%
\varphi(x+(p-c)h-U_{1},c)\right]  -\varphi(x,c).
\end{array}
\]
So, dividing by $h$ and taking $h\rightarrow0^{+}$, we get%

\[
\mathcal{L}(\varphi)(x,c)\leq0
\]
and so it is a viscosity supersolution at $(x,c)$.

Let us prove now that $V$ is a viscosity subsolution in $(0,\infty
)\times\lbrack0,\overline{c})$. Arguing by contradiction, we assume that $V $
is not a subsolution of (\ref{HJB equation}) at $\left(  x,c\right)
\in(0,\infty)\times\lbrack0,\overline{c})$, then there exist $\varepsilon>0$,
$0<h<\min\left\{  x/2,\overline{c}-c\right\}  $ and a continuously
differentiable function $\psi$ with $\psi(x,c)=V(x,c)$ such that $\psi\geq V$,%

\begin{equation}
\max\{\mathcal{L}(\psi)(y,d),\psi_{c}(y,d)\}\leq-q\varepsilon<0
\label{Desig1aprima}%
\end{equation}
for $\left(  y,d\right)  \in$ $[x-h,x+h]\times\lbrack c,c+h]$ and%

\begin{equation}
V(y,d)\leq\psi(y,d)-\varepsilon\label{Desig2a}%
\end{equation}
for $\left(  y,d\right)  \notin\lbrack x-h,x+h]\times\lbrack c,c+h]$.

Consider the controlled risk process $X_{t}$ corresponding to an admissible
strategy $C\in\Pi_{x,c,\overline{c}}$ and define
\[
\tau^{\ast}=\inf\{t>0:\text{ }\left(  X_{t},C_{t}\right)  \notin\lbrack
x-h,x+h]\times\lbrack c,c+h]\}\text{.}%
\]
Since $C_{t}$ is non-decreasing and right-continuous, it can be written as
\begin{equation}
C_{t}=c+\int\nolimits_{0}^{t}dC_{s}^{co}+\sum_{\substack{C_{s}\neq C_{s^{-}}
\\0\leq s\leq t}}(C_{s}-C_{s^{-}}) \label{Lt para ITO}%
\end{equation}
where $C_{s}^{co}$ is a continuous and non-decreasing function.

Take a non-negative continuously differentiable function $\psi(x,c)$\ in
$(0,\infty)\times\lbrack0,\overline{c}]$. Since the function $e^{-qt}%
\psi(x,c)$ is continuously differentiable, using the expression
(\ref{Lt para ITO}) and the change of variables formula for finite variation
processes (see for instance \cite{Protter}), we can write
\begin{equation}%
\begin{array}
[c]{l}%
\psi(X_{\tau^{\ast}}^{C},C_{^{\tau^{\ast-}}})e^{-q\tau^{\ast}}-\psi(x,c)\\%
\begin{array}
[c]{ll}%
= & \int\nolimits_{0}^{\tau^{\ast}}e^{-qs}\psi_{x}(X_{s^{-}}^{C},C_{s^{-}%
})(p-C_{s^{-}})ds+\int\nolimits_{0}^{\tau^{\ast}}e^{-qs}\psi_{c}(X_{s^{-}}%
^{C},C_{s^{-}})dC_{s}^{co}\\
& +\sum_{\substack{C_{s}\neq C_{s^{-}}\\0\leq s<\tau^{\ast}}}e^{-qs}%
(C_{s}-C_{s^{-}})\psi_{c}(X_{s^{-}}^{C},c_{s})\\
& +\sum\limits_{\tau_{i}<\tau^{\ast}}\left(  \psi(X_{s^{-}}^{C}-U_{i}%
,C_{s^{-}})-\psi(X_{s^{-}}^{C},C_{s^{-}})\right)  e^{-qs}-q\int\nolimits_{0}%
^{\tau^{\ast}}\psi(X_{s^{-}}^{C},C_{s^{-}})e^{-qs}ds,
\end{array}
\end{array}
\label{Paso 1}%
\end{equation}
where $c_{s}\in(C_{s^{-}},C_{s})$. We have that%
\[%
\begin{array}
[c]{cl}%
M_{t}= & \sum\limits_{\tau_{i}\leq\tau}\left(  \psi(X_{s^{-}}^{C}%
-U_{i},C_{s^{-}})-\psi(X_{s^{-}}^{C},C_{s^{-}})\right)  e^{-qs}\\
& -\beta\int\nolimits_{0}^{t}e^{-qs}\int\nolimits_{0}^{\infty}\left(
\psi(X_{s^{-}}^{C}-\alpha,C_{s^{-}})-\psi(X_{s^{-}}^{C},C_{s^{-}})\right)
dF(\alpha)ds
\end{array}
\]
is a martingale with zero expectation.\ Hence, from (\ref{Desig1aprima}), we
can write%
\[%
\begin{array}
[c]{l}%
e^{-q\tau^{\ast}}\psi(X_{\tau^{\ast}},C_{\tau^{\ast-}})-\psi(x,c)\\%
\begin{array}
[c]{ll}%
= & \int\nolimits_{0}^{\tau^{\ast}}e^{-qs}\mathcal{L}_{C_{s^{-}}}%
(\psi)(X_{s^{-}}^{C},C_{s^{-}})ds+M_{\tau^{\ast-}}-\int\nolimits_{0}%
^{\tau^{\ast}}e^{-qs}C_{s^{-}}ds\\
& +\int\nolimits_{0}^{\tau^{\ast}}e^{-qs}\psi_{c}(X_{s^{-}}^{C},C_{s^{-}%
})dC_{s}^{c}+\sum_{\substack{C_{s}\neq C_{s^{-}}\\0\leq s<\tau^{\ast}}%
}e^{-qs}(C_{s}-C_{s^{-}})\psi_{c}(X_{s^{-}}^{C},c_{s})\\
\leq & \int\nolimits_{0}^{\tau^{\ast}}e^{-qs}(-q\varepsilon)ds+M_{\tau^{\ast
-}}-\int\nolimits_{0}^{\tau^{\ast}}e^{-qs}C_{s^{-}}ds\\
& +\int\nolimits_{0}^{\tau^{\ast}}e^{-qs}(-q\varepsilon)dC_{s}^{c}%
+\sum_{\substack{C_{s}\neq C_{s^{-}}\\0\leq s<\tau^{\ast}}}e^{-qs}%
(C_{s}-C_{s^{-}})(-q\varepsilon)\\
= & \varepsilon\left(  e^{-q\tau^{\ast}}-1\right)  +M_{\tau^{\ast-}}%
-\int\nolimits_{0}^{\tau^{\ast}}e^{-qs}C_{s^{-}}ds-q\varepsilon\left(
\int\nolimits_{0}^{\tau^{\ast}}e^{-qs}dC_{s}\right)  .
\end{array}
\end{array}
\]

So, taking expectation and using that $V$ is non-increasing in $c$, we obtain
from (\ref{Desig2a}) that%
\[%
\begin{array}
[c]{l}%
\mathbb{E}\left[  e^{-q\tau^{\ast}}V(X_{\tau^{\ast}},C)_{\tau^{\ast}}\right]
\\%
\begin{array}
[c]{cl}%
= & \mathbb{E}\left[  e^{-q\tau^{\ast}}(V(X_{\tau^{\ast}},C_{\tau^{\ast}%
})-V(X_{\tau^{\ast}},C_{\tau^{\ast-}}))\right]  +\mathbb{E}\left[
e^{-q\tau^{\ast}}V(X_{\tau^{\ast}},C_{\tau^{\ast-}})\right]  \\
\leq & \mathbb{E}\left[  \psi(x,c)-e^{-q\tau^{\ast}}\varepsilon\right]
+\mathbb{E}\left[  \psi(X_{\tau^{\ast}},C_{\tau^{\ast-}})e^{-q\tau^{\ast}%
}-\psi(x,c)\right]  \\
\leq & \psi(x,c)-\varepsilon\mathbb{E}\left[  e^{-q\tau^{\ast}}\right]
-\left(  \varepsilon\mathbb{E}\left[  1-e^{-q\tau^{\ast}}\right]
+q\varepsilon\mathbb{E}\left[  \int\nolimits_{0}^{\tau^{\ast}}e^{-qs}%
dC_{s}\right]  \right)  \ -\mathbb{E}\left[  \int\nolimits_{0}^{\tau^{\ast}%
}e^{-qs}C_{s^{-}}ds\right]  \\
\leq & \psi(x,c)-\varepsilon\mathbb{E}\left[  e^{-q\tau^{\ast}}\right]
-\varepsilon\mathbb{E}\left[  1-e^{-q\tau^{\ast}}\right]  -\mathbb{E}\left[
\int\nolimits_{0}^{\tau^{\ast}}e^{-qs}C_{s^{-}}ds\right]  \ \\
= & \psi(x,c)-\varepsilon-\mathbb{E}(\int\nolimits_{0}^{\tau^{\ast}}%
e^{-qs}C_{s^{-}}ds).
\end{array}
\end{array}
\]
Hence, using the dynamic programming principle (\ref{DPP}), we have that
\[
V(x,c)=\sup\limits_{C\in\Pi_{x,c,\overline{c}}}\mathbb{E}\left(
\int\nolimits_{0}^{\tau^{\ast}}e^{-qs}C_{s^{-}}ds+e^{-c\tau^{\ast}}%
V(X_{\tau^{\ast}}^{C},C_{\tau^{\ast}})\right)  \leq\psi(x,c)-\varepsilon.
\]
but this is a contradiction in view of the assumption $V(x,c)=\psi(x,c)$.
\hfill$\blacksquare$\newline

$V(x,\overline{c})$ corresponds to the value function of the strategy that
pays dividends at constant rate $\overline{c}$, so the following lemma is a
standard one-dimensional result.

\begin{lemma}
\label{V en cbarra}The optimal value function $V(x,\overline{c})$ is a
solution of $\mathcal{L}(V)(x,\overline{c})=0$ for $x>0$, where $\mathcal{L}$
is the operator defined in (\ref{Operador L}).
\end{lemma}

We now show that $V$ satisfies a boundary condition as $x$ goes to infinity.

\begin{proposition}
\label{lim l0/c}The optimal value function $V$ satisfies $\lim_{x\rightarrow
\infty}V(x,c)=\overline{c}/q$ for all $c\in\lbrack0,\overline{c}].$
\end{proposition}

\emph{Proof.} We first prove the result for $c=\overline{c}.$ Since
$V(\cdot,\overline{c}):\left[  0,\infty\right)  \rightarrow\mathbb{R}$ is
bounded, non-decreasing and Lipschitz, the set $\widetilde{\mathcal{D}}$ of
points where $V(\cdot,\overline{c})$ is differentiable has full measure in
$[0,\infty)$. Let us prove that there exists a sequence $x_{n}\rightarrow
\infty$ such that $x_{n}\in\widetilde{\mathcal{D}},$ $V_{x}(x_{n},\overline
{c})\rightarrow0$. Given any $\varepsilon>0$, if $V_{x}(x,\overline{c}%
)\geq\varepsilon$ for all $x\in\widetilde{\mathcal{D}}\cap\lbrack n,\infty)$
with $n\in\mathbb{N}$, then $V(x,\overline{c})$ cannot be bounded and this is
a contradiction, so such a sequence exists. We obtain%
\[%
\begin{array}
[c]{lll}%
0 & = & \lim\limits_{n\rightarrow\infty}\mathcal{L}(V)(x_{n},\overline{c})\\
& = & \overline{c}+(p-\overline{c})\lim\limits_{n\rightarrow\infty}V_{x}%
(x_{n},\overline{c})-(q+\beta)\lim_{x\rightarrow\infty}V(x_{n},\overline
{c})+\beta\int\nolimits_{0}^{\infty}\lim_{x\rightarrow\infty}V(x,\overline
{c})dF(\alpha)\\
& = & \overline{c}-q\lim_{x\rightarrow\infty}V(x,\overline{c}).~
\end{array}
\]
Finally, for $c\in\lbrack0,\overline{c})$, since $V(x,c)\leq\overline{c}/q$
but also $V(x,c)\geq V(x,\overline{c})$ for all $c\in\left[  0,\overline
{c}\right]$, so that we obtain the result.\hfill$\blacksquare$

We now give the comparison result for viscosity solutions.

\begin{lemma}
\label{Lema para Unicidad} In the case $\overline{c}\leq p$ consider any
interval $[c_{1},c_{2}]\subset\lbrack0,\overline{c}]$ and in the case
$\overline{c}>p$ consider any interval $[c_{1},c_{2}]\subset\lbrack0,p]$ or
any interval $[c_{1},c_{2}]\subset(p,\overline{c}]$. Let us assume that (i)
$\underline{u}$ is a viscosity subsolution and $\overline{u}$ is a viscosity
supersolution of the HJB equation (\ref{HJB equation}) for all $x>0$ and for
all $c\in(c_{1},c_{2})$, (ii) $\underline{u}$ and $\overline{u}$ are
non-decreasing in the variable $x$ and Lipschitz in $[0,\infty)\times\lbrack
c_{1},c_{2}]$, (iii) $\lim_{x\rightarrow\infty}\underline{u}(x,c)=\lim
_{x\rightarrow\infty}\overline{u}(x,c)=L>0$ and (iv) $\underline{u}%
(x,c_{2})\leq\overline{u}(x,c_{2})$ for all $x\geq0$; then $\underline{u}%
\leq\overline{u}$ in $[0,\infty)\times\lbrack c_{1},c_{2}].$
\end{lemma}

\emph{Proof.} The proof of this lemma is a two-dimensional generalization of
Proposition 4.2 of Azcue and Muler \cite{AM Bounded 2012}. We use in this
proof an equivalent formulation of viscosity solution, see for example Sayah
\cite{Sayah} and Benth, Karlsen and Reikvam \cite{Benth}: Let us define the
operators%
\[%
\begin{array}
[c]{l}%
\mathcal{L}(u,\psi)(x,c)=c+(p-c)\psi_{x}(x,c)-(q+\beta)u(x,c)+\beta
\int\nolimits_{0}^{x}u(x-\alpha,c)dF(\alpha)\text{ and}\\
\overline{\mathcal{L}}(\psi)(x,c)=\psi_{c}(x,c).
\end{array}
\]
A locally Lipschitz function $\overline{u}$\ $:[0,\infty)\times\lbrack
c_{1},c_{2}]\rightarrow\mathbb{R}$\ is a viscosity supersolution of
(\ref{HJB equation}) at $(x,c)\in(0,\infty)\times(c_{1},c_{2})$\ if any test
function $\varphi$ for supersolution at $(x,c)$ satisfies
\[
\max\{\mathcal{L}(\overline{u},\varphi)(x,c),\overline{\mathcal{L}}%
(\varphi)(x,c)\}\leq0\text{,}%
\]
and a locally Lipschitz function $\underline{u}:[0,\infty)\times\lbrack
c_{1},c_{2}]\rightarrow\mathbb{R}$\ is a viscosity subsolution\ of
(\ref{HJB equation}) at $(x,c)\in(0,\infty)\times(c_{1},c_{2})$\ if any test
function $\psi$ for subsolution at $(x,c)$ satisfies
\[
\max\{\mathcal{L}(\underline{u},\psi)(x,c),\overline{\mathcal{L}}%
(\psi)(x,c)\}\geq0.
\]

Suppose that there is a point $(x_{0},c_{0})\in\lbrack0,\infty)\times
(c_{1},c_{2})$ such that $\underline{u}(x_{0},c_{0})-\overline{u}(x_{0}%
,c_{0})>0$. Let us define $h(c)=1+\eta e^{-\frac{c}{c_{2}}}$ with
$\eta=(\underline{u}(x_{0},c_{0})-\overline{u}(x_{0},c_{0}))/(2\overline
{u}(x_{0},c_{0}))$,$\ $and
\[
\overline{u}^{s}(x,c)=sh(c)\overline{u}(x,c)
\]
for any $s>1$. We have that $\varphi$ is a test function for supersolution of
$\overline{u}$ at $(x,c)$ if and only if $\varphi^{s}=sh(c)\varphi$ is a test
function for supersolution of $\overline{u}^{s}$ at $(x,c)$. We have
\begin{equation}
\mathcal{L}(\overline{u}^{s},~\varphi^{s})(x,c)=sh(c)\mathcal{L}(\overline
{u},\varphi)(x,c)+c(1-sh(c))<0,\label{Desigualdad L1}%
\end{equation}
and%
\begin{equation}
\overline{\mathcal{L}}(\varphi^{s})(x,c)\leq-\frac{s}{c_{2}}\varphi
(x,c)e^{-\frac{c}{c_{2}}}<0\label{Desigualdad L2}%
\end{equation}
for $\varphi(x,c)>0.$ Take $s_{0}>1$ such that $\underline{u}(x_{0}%
,c_{0})-\overline{u}^{s_{0}}(x_{0},c_{0})>0$. We define%
\begin{equation}
M=\sup\limits_{x\geq0,c_{1}\leq c\leq c_{2}}\left(  \underline{u}%
(x,c)-\overline{u}^{s_{0}}(x,c)\right)  .\label{Definicion de M}%
\end{equation}

Since $\lim_{x\rightarrow\infty}\underline{u}(x,c)=\lim_{x\rightarrow\infty
}\overline{u}(x,c)=L$, we have that there exists a $b>x_{0}$ such that%

\begin{equation}
\sup\limits_{x\geq0,c_{1}\leq c,d\leq c_{2}}\underline{u}(x,c)-\overline
{u}^{s_{0}}(x,d)<0\text{ for }x\geq b. \label{desborde}%
\end{equation}
We obtain from (\ref{desborde}) that%

\begin{equation}
0<\underline{u}(x_{0},c_{0})-\overline{u}^{s_{0}}(x_{0},c_{0})\leq
M:=\max\limits_{x\in\left[  0,b\right]  ,c_{1}\leq c\leq c_{2}}\left(
\underline{u}(x,c)-\overline{u}^{s_{0}}(x,c)\right)  . \label{desmax}%
\end{equation}
Call $\left(  x^{\ast},c^{\ast}\right)  :=\arg\max\limits_{x\in\left[
0,b\right]  ,c_{1}\leq c\leq c_{2}}\left(  \underline{u}(x,c)-\overline
{u}^{s_{0}}(x,c)\right)  $. Since $\underline{u}$ and $\overline{u}^{s_{0}}$
are locally Lipschitz, there exists a constant $m>0$ such that%
\begin{equation}
\left\vert \underline{u}(x,c)-\underline{u}(y,d)\right\vert \leq m\left\Vert
(x-y,c-d)\right\Vert _{2}\text{ and }\left\vert \overline{u}^{s_{0}%
}(x,c)-\overline{u}^{s_{0}}(y,d)\right\vert \leq m\left\Vert
(x-y,c-d)\right\Vert _{2}\text{ } \label{acotacion-con-m}%
\end{equation}
for $0\leq x_{2}\leq x_{1}\leq b$. Consider the set%
\[
A=\left\{  \left(  x,y,c,d\right)  :0\leq x\leq y\leq b,c_{1}\leq\ c\leq
c_{2},c_{1}\leq d\leq c_{2}\right\}
\]
and, for all $\lambda>0$, the functions%
\begin{equation}%
\begin{array}
[c]{l}%
\Phi^{\lambda}\left(  x,y,c,d\right)  =\dfrac{\lambda}{2}\left(  x-y\right)
^{2}+\dfrac{\lambda}{2}\left(  c-d\right)  ^{2}+\frac{2m}{\lambda^{2}\left(
y-x\right)  +\lambda},\\
\Sigma^{\lambda}\left(  x,y,c,d\right)  =\underline{u}(x,c)-\overline
{u}^{s_{0}}(y,d)-\Phi^{\lambda}\left(  x,y,c,d\right)  .
\end{array}
\label{def-sigma-definitiva-por-hoy}%
\end{equation}
We have that the partial derivatives are%
\[%
\begin{array}
[c]{l}%
\Phi_{x}^{\lambda}\left(  x,y,c,d\right)  =\lambda\left(  x-y\right)
+\frac{2m}{\left(  \lambda\left(  y-x\right)  +1\right)  ^{2}},~\Phi
_{y}^{\lambda}\left(  x,y,c,d\right)  =-\lambda\left(  x-y\right)  -\frac
{2m}{\left(  \lambda\left(  y-x\right)  +1\right)  ^{2}};\\
\Phi_{c}^{\lambda}\left(  x,y,c,d\right)  =\lambda\left(  c-d\right)  \text{
and }\Phi_{d}^{\lambda}\left(  x,y,c,d\right)  =-\lambda\left(  c-d\right)  .
\end{array}
\]
Calling $M^{\lambda}=\max\limits_{A}\Sigma^{\lambda}$ and $\left(  x_{\lambda
},y_{\lambda},c_{\lambda},d_{\lambda}\right)  =\arg\max\limits_{A}%
\Sigma^{\lambda}$, we obtain that $M^{\lambda}\geq\Sigma^{\lambda}(x^{\ast
},x^{\ast},c^{\ast},c^{\ast})=M$ and so%

\begin{equation}
\liminf\limits_{\lambda\rightarrow\infty}M^{\lambda}\geq M.
\label{liminf-mlambda}%
\end{equation}

We show that there exists $\lambda_{0}$ large enough such that if $\lambda
\geq\lambda_{0}$, then $\left(  x_{\lambda},y_{\lambda},c_{\lambda}%
,d_{\lambda}\right)  $ $\notin\partial A$. The maximum is not achieved on the
boundary $y=x$ because%
\begin{equation}
\lim\inf_{h\searrow0}\tfrac{\Sigma^{\lambda}\left(  x,x+h,c,d\right)
-\Sigma^{\lambda}\left(  x,x,c,d\right)  }{h}=\lim\inf_{h\searrow0}%
\tfrac{\overline{u}^{s_{0}}(x,d)-\overline{u}^{s_{0}}(x+h,d)}{h}-\Phi
_{y}^{\lambda}\left(  x,x,c,d\right)  \geq m>0. \label{Borde x=y}%
\end{equation}
Let us see now that the maximum is also not achieved on the boundary $x=0.$
Since $\Sigma^{\lambda}$ is continuous and locally Lipschitz, by
(\ref{Borde x=y}) there exists an open set
\[
O_{1}\supset\left\{  \left(  0,0\right)  \right\}  \times\lbrack c_{1}%
,c_{2}]\times\lbrack c_{1},c_{2}],
\]
where $\Sigma^{\lambda}$ does not achieve the maximum. Correspondingly, there
exists an $\varepsilon>0$ such that the maximum of $\Sigma^{\lambda}$ is not
achieved at the points $(0,y,c,d)$ with $0\leq y<\varepsilon.$ Moreover, since
$\underline{u}$ is a non-decreasing function in $x$, we have from
(\ref{desborde}) and (\ref{def-sigma-definitiva-por-hoy}) that%

\begin{equation}%
\begin{array}
[c]{lll}%
\lim\inf_{h\searrow0}\frac{\Sigma^{\lambda}\left(  h,y,c,d\right)
-\Sigma^{\lambda}\left(  0,y,c,d\right)  }{h} & = & \lim\inf_{h\searrow0}%
\frac{\underline{u}(h,c)-\underline{u}(0,c)}{h}-\Phi_{x}^{\lambda}\left(
0,y,c,d\right) \\
& \geq & \lambda y-\frac{2m}{\left(  \lambda y+1\right)  ^{2}}>0
\end{array}
\label{Borde x=0}%
\end{equation}
for $\lambda$ large enough if $y>\varepsilon>0$; so the maximum is not
achieved on the boundary $x=0$. With similar arguments, it can be proved that
the maximum is not achieved on the boundaries $y=b,$ $c=c_{1},~c=c_{2}%
,~d=c_{1}\ $and $d=c_{2}$.

\bigskip Since $\Sigma^{\lambda}\left(  x,y,c,d\right)  =\underline
{u}(x,c)-\overline{u}^{s_{0}}(y,d)-\Phi^{\lambda}\left(  x,y,c,d\right)  $
reaches the maximum in $\left(  x_{\lambda},y_{\lambda},c_{\lambda}%
,d_{\lambda}\right)  \ $ in the interior of the set $A,$ the function
\[
\psi(x,c)=\Phi^{\lambda}\left(  x,y_{\lambda},c,d_{\lambda}\right)
-\Phi^{\lambda}\left(  x_{\lambda},y_{\lambda},c_{\lambda},d_{\lambda}\right)
+\underline{u}\left(  x_{\lambda},c_{\lambda}\right)
\]
is a test for subsolution for $\underline{u}$ at $\left(  x_{\lambda
},c_{\lambda}\right)  $, and so%
\begin{equation}
\max\{\mathcal{L}(\underline{u},\psi)(x_{\lambda},c_{\lambda}),\overline
{\mathcal{L}}(\psi)(x_{\lambda},c_{\lambda})\}\geq0. \label{subxalfa1}%
\end{equation}
Furthermore,
\[
\varphi^{s_{0}}(y,d)=-\Phi^{\lambda}\left(  x_{\lambda},y,c_{\lambda
},d\right)  +\Phi^{\lambda}\left(  x_{\lambda},y_{\lambda},c_{\lambda
},d_{\lambda}\right)  +\overline{u}^{s_{0}}\left(  y_{\lambda},d_{\lambda
}\right)
\]
is a test for supersolution for $\overline{u}^{s_{0}}$ at $\left(  y_{\lambda
},d_{\lambda}\right)  $ and so%
\begin{equation}
\max\{\mathcal{L}(\overline{u}^{s_{0}},\varphi^{s_{0}})\left(  y_{\lambda
},d_{\lambda}\right)  ,\overline{\mathcal{L}}(\varphi^{s_{0}})\left(
y_{\lambda},d_{\lambda}\right)  \}\leq0. \label{superyalfa}%
\end{equation}
Since
\[
\overline{\mathcal{L}}(\varphi^{s})(y_{\lambda},d_{\lambda})\leq-\frac{s_{0}%
}{c_{2}}\varphi(y_{\lambda},d_{\lambda})e^{-\frac{c}{c_{2}}}<0
\]
(because $y_{\lambda}>0)$ and $\overline{\mathcal{L}}(\psi)(x_{\lambda
},c_{\lambda})$ $=\overline{\mathcal{L}}(\varphi^{s})(y_{\lambda},d_{\lambda
})<0$, we have from (\ref{subxalfa1}) that%

\begin{equation}
\mathcal{L}(\underline{u},\psi)(x_{\lambda},c_{\lambda})\geq0.
\label{subxalfa2}%
\end{equation}

Therefore, from (\ref{superyalfa}), (\ref{subxalfa1}) and $\psi_{x}%
(x_{\lambda},c_{\lambda})=\varphi_{x}^{s}(y_{\lambda},d_{\lambda}),$ we get%

\begin{equation}%
\begin{array}
[c]{l}%
\mathcal{L}(\overline{u}^{s_{0}},\varphi^{s_{0}})\left(  y_{\lambda
},d_{\lambda}\right)  -\mathcal{L}(\underline{u},\psi)(x_{\lambda},c_{\lambda
})\\%
\begin{array}
[c]{l}%
=(q+\beta)\left(  \underline{u}(x_{\lambda},c_{\lambda})-\overline{u}^{s_{0}%
}(y_{\lambda},d_{\lambda})\right) \\
\leq\beta\left(  \int\nolimits_{0}^{x_{\lambda}}\underline{u}(x_{\lambda
}-\alpha,c_{\lambda})dF(\alpha)-\int\nolimits_{0}^{y_{\lambda}}\overline
{u}^{s_{0}}(y_{\lambda}-\alpha,d_{\lambda})dF(\alpha)\right)  .
\end{array}
\end{array}
\label{desdifsup1}%
\end{equation}

Using the inequality%

\[
\Sigma^{\lambda}\left(  x_{\lambda},x_{\lambda},c_{\lambda},c_{\lambda
}\right)  +\Sigma^{\lambda}\left(  y_{\lambda},y_{\lambda},d_{\lambda
},d_{\lambda}\right)  \leq2\Sigma^{\lambda}\left(  x_{\lambda},y_{\lambda
},c_{\lambda},d_{\lambda}\right)
\]
we obtain that%
\[
\lambda\left(  x_{\lambda}-y_{\lambda}\right)  ^{2}+\lambda\left(  c_{\lambda
}-d_{\lambda}\right)  ^{2}\leq\underline{u}(x_{\lambda},c_{\lambda
})-\underline{u}(y_{\lambda},d_{\lambda})+\overline{u}^{s_{0}}(x_{\lambda
},c_{\lambda})-\overline{u}^{s_{0}}(y_{\lambda},d_{\lambda})+4m(y_{\lambda
}-x_{\lambda}),
\]
which together with (\ref{acotacion-con-m}) gives
\begin{equation}
\lambda\left\Vert (x_{\lambda}-y_{\lambda},c_{\lambda}-d_{\lambda})\right\Vert
_{2}^{2}\leq6m\left\Vert (x_{\lambda}-y_{\lambda},c_{\lambda}-d_{\lambda
})\right\Vert _{2}. \label{desxalfa-yalfac}%
\end{equation}
We can find a sequence $\lambda_{n}\rightarrow\infty$ such that $\left(
x_{\lambda_{n}},y_{\lambda_{n}},c_{\lambda_{n}},d_{\lambda_{n}}\right)
\rightarrow\left(  \widehat{x},\widehat{y},\widehat{c},\widehat{d}\right)  \in
A$. From (\ref{desxalfa-yalfac}), we get that $\left\Vert (x_{\lambda_{n}%
}-y_{\lambda_{n}},c_{\lambda_{n}}-d_{\lambda_{n}})\right\Vert _{2}\leq
\dfrac{6m}{\lambda_{n}}$, which gives $\widehat{x}=\widehat{y}$ and
$\widehat{c}=\widehat{d}$. Using that $y_{\lambda_{n}}\geq x_{\lambda_{n}}$
for all $n$, we obtain from (\ref{desdifsup1}) that%
\begin{equation}
(q+\beta)\left(  \underline{u}(\widehat{x},\widehat{c})-\overline{u}^{s_{0}%
}(\widehat{x},\widehat{c})\right)  \leq\beta\left(  \int\nolimits_{0}%
^{\widehat{x}}\left(  \underline{u}(\widehat{x}-\alpha,\widehat{c}%
)-\overline{u}^{s_{0}}(\widehat{x}-\alpha,\widehat{c})\right)  dF(\alpha
)\right)  \leq\beta M. \label{Desigualdad-Integral-Uriel}%
\end{equation}
From (\ref{desxalfa-yalfac}) we get that $\lim\limits_{n\rightarrow\infty
}\lambda_{n}\left\Vert (x_{\lambda_{n}}-y_{\lambda_{n}},c_{\lambda_{n}%
}-d_{\lambda_{n}})\right\Vert _{2}^{2}=0$, hence from (\ref{liminf-mlambda})
and (\ref{Desigualdad-Integral-Uriel}) we obtain%
\begin{align*}
M  &  \leq\liminf\limits_{\lambda\rightarrow\infty}M_{\lambda}\leq
\lim\limits_{_{n\rightarrow\infty}}M_{\lambda_{n}}=\lim\limits_{_{n\rightarrow
\infty}}\Sigma^{\lambda_{x}}(x_{\lambda_{n}},y_{\lambda_{n}},c_{\lambda_{n}%
},d_{\lambda_{n}})=\underline{u}(\widehat{x},\widehat{c})-\overline{u}^{s_{0}%
}(\widehat{x},\widehat{c})\\
&  \leq\frac{\beta}{q+\beta}\int\nolimits_{0}^{\widehat{x}}\left(
\underline{u}(\widehat{x}-\alpha,\widehat{c})-\overline{u}^{s_{0}}(\widehat
{x}-\alpha,\widehat{c})\right)  dF(\alpha)\leq\dfrac{\beta}{q+\beta}M.
\end{align*}
This is a contradiction, which establishes the result.\hfill$\blacksquare
$\newline

As a consequence of the previous lemma, we have the following two propositions
concerning uniqueness for the cases $\overline{c}\leq p$ and
$\overline{c}>p$. In the case $\overline{c}\leq p$ the uniqueness is a direct
consequence of Lemmas \ref{Lema para Unicidad} and \ref{lim l0/c} together
with Proposition \ref{Proposicion Viscosidad}.

\begin{proposition}
If $\overline{c}\leq p$, the optimal value function $V$ is the unique function
non-decreasing in $x$ that is a viscosity solution of (\ref{HJB equation}) in
$(0,\infty)\times\lbrack0,\overline{c})$ with limit $\overline{c}/q$ as
$x\rightarrow\infty$.
\end{proposition}

In the case $\overline{c}>p$ the uniqueness is a direct consequence of Lemmas
\ref{Lema para Unicidad} and \ref{lim l0/c} together with Propositions
\ref{Proposicion Viscosidad} and \ref{Continuidad en p en la variable c}.

\begin{proposition}
If $\overline{c}>p$, the optimal value function $V$ is the unique continuous
function non-decreasing in $x$ that has limit $\overline{c}/q$ as
$x\rightarrow\infty$ for all $c\in\left[  0,\overline{c}\right]  $ and that is
a viscosity solution of (\ref{HJB equation}) both in $[0,\infty)\times
\lbrack0,p]$ and in $[0,\infty)\times\lbrack c_{1},\overline{c})$ for any
$c_{1}\in(p,\overline{c})$.
\end{proposition}

From Definition \ref{optimal value function}, Lemma \ref{Lema para Unicidad},
and \ref{lim l0/c} together with Proposition \ref{Proposicion Viscosidad}, we
also get the following verification theorem that will be used in the next section.

\begin{theorem}
\label{verification result} In the case $\overline{c}\leq p$ consider any
interval $[c_{1},c_{2}]\subset\lbrack0,\overline{c}]$ and in the case
$\overline{c}>p$ consider any interval $[c_{1},c_{2}]\subset\lbrack0,p]$ or
any interval $[c_{1},c_{2}]\subset(p,\overline{c}]$. Consider a family of
strategies $\left\{  C_{x,c}\in\Pi_{x,c,\overline{c}}:(x,c)\in\lbrack
0,\infty)\times\lbrack c_{1},c_{2}]\right\}  $. If the function
$W(x,c):=J(x;C_{x,c})$ is a viscosity supersolution of the HJB equation
(\ref{HJB equation}) in $(0,\infty)\times(c_{1},c_{2})$ with $\lim
_{x\rightarrow\infty}W(x,c)=$ $\overline{c}/q$, then $W$ is the optimal value
function $V$. Also, if for each $k\geq1$ there exists a family of strategies
$\left\{  C_{x,c}^{k}\in\Pi_{x,c,\overline{c}}:(x,c)\in\lbrack0,\infty
)\times\lbrack c_{1},c_{2}]\right\}  $ such that $W(x,c):=\lim_{k\rightarrow
\infty}J(x;C_{x,c}^{k})$ is a viscosity supersolution of the HJB equation
(\ref{HJB equation}) in $(0,\infty)\times(c_{1},c_{2})$ with $\lim
_{x\rightarrow\infty}W(x,c)=$ $\overline{c}/q$, then $W$ is the optimal value
function $V$.
\end{theorem}

\section{Finite ratcheting strategies}

\label{sec5}

In this section we introduce ratcheting strategies when only a finite number
$N$ of dividend rates are possible and find the optimal value
function in this restricted setting. This optimization problem is no longer
two-dimensional and can be reduced to $N$ one-dimensional obstacle problems.
We also show that there exists an optimal finite ratcheting strategy and we
construct it recursively. In Section \ref{sec6} we will then use the optimal
value function of this restricted setting to approximate the optimal value
function $V$ for the general case.

Consider a finite set $\mathcal{G=\{}c_{1},c_{2},\ldots c_{N}\mathcal{\}}$ in
the interval $[0,\overline{c}]$ with $c_{k}<c_{k+1}$ and $c_{N}=\overline{c}$.
The task is then to find the optimal value function among the ratcheting
strategies with dividend rates in $\mathcal{G}$. To that end, let us define
the family of admissible strategies $\Pi_{x,c,\overline{c}}^{\mathcal{G}%
}\subset\Pi_{x,c,\overline{c}}$ as
\[
\Pi_{x,c,\overline{c}}^{\mathcal{G}}=\{C\in\Pi_{x,c,\overline{c}}\text{ such
that }\operatorname{Im}(C)\subset\mathcal{G}\}\text{.}%
\]
and the optimal value function within the restricted class as
\begin{equation}
V^{\mathcal{G}}(x,c)=\sup_{C\in\Pi_{x,c,\overline{c}}^{\mathcal{G}}}J(x;C).
\label{Definicion VN}%
\end{equation}
By definition, $V^{\mathcal{G}}(x,c)=V^{\mathcal{G}}(x,\widetilde{c})$ where
$\widetilde{c}=\min\{c_{k}\in\mathcal{G}:c_{k}\geq c\}$ and $V^{\mathcal{G}%
}(x,c)\leq V(x,c)$ for all finite sets $\mathcal{G}$.\newline

Let us first state some basic properties of $V^{\mathcal{G}}$. These
properties mirror the properties of the optimal value function $V.$ The proofs
are the same as those of Propositions \ref{Monotone Optimal Value Function},
\ref{Proposition Global Lipschitz zone},
\ref{Proposition Lipschitz cmax mayor a p_Alternativa} and \ref{lim l0/c} but
considering admissible strategies in the set $\Pi_{x,c,\overline{c}%
}^{\mathcal{G}}$ instead of $\Pi_{x,c,\overline{c}}.$

\begin{proposition}
\label{Lipchitz de Vn}

\begin{enumerate}
\item $V^{\mathcal{G}}(x,c)$ is non-increasing in $c$ with $V^{\mathcal{G}%
}(x,\overline{c})=V(x,\overline{c})$, and non-decreasing in $x$ with
$\lim_{x\rightarrow\infty}$ $V^{\mathcal{G}}(x,c)=\overline{c}/q$.

\item There exists a constant $K_{1}>0$ such that
\[
0\leq V^{\mathcal{G}}(x_{2},c_{k})-V^{\mathcal{G}}(x_{1},c_{l})\leq
K_{1}\left[  \left(  x_{2}-x_{1}\right)  +\left(  c_{l}-c_{k}\right)  \right]
\]
for all $0\leq x_{1}\leq x_{2},$ and $c_{k}\in\mathcal{G},c_{l}\in\mathcal{G}$
with $c_{k}\leq c_{l}\leq\min\left\{  \overline{c},p\right\}  .$

\item Assume that $\overline{c}>p$, then there exist constants $K_{2\text{ }%
}>0$ and $K_{3}>0$ such that%
\[
0\leq V^{\mathcal{G}}(x_{2},c_{k})-V^{\mathcal{G}}(x_{1},c_{l})\leq\left[
K_{2}+\frac{K_{3}}{c_{k}-p}\right]  \left(  x_{2}-x_{1}\right)  +\left[
K_{2}+\frac{K_{3}x_{2}}{\left(  c_{k}-p\right)  ^{2}}\right]  \left(
c_{l}-c_{k}\right)
\]
for all $0\leq x_{1}\leq x_{2},$ and $c_{k}\in\mathcal{G},c_{l}\in\mathcal{G}$
with $p<c_{k}\leq c_{l}\leq\overline{c}.$
\end{enumerate}
\end{proposition}

\bigskip From the problem definition, for any given finite set $\mathcal{G=\{}%
c_{1},c_{2},\ldots c_{N}\mathcal{\}}$ in the interval $[0,\overline{c}]$ with
$c_{k}<c_{k+1}$ and $c_{N}=\overline{c}$, we have that%
\[
V^{\mathcal{G}}(x,c_{N})=V^{\mathcal{G}}(x,\overline{c})=\mathbb{E}\left[
\int_{0}^{\tau}e^{-qs}\,\overline{c}\,ds\right]  .
\]
We can now describe $V^{\mathcal{G}}(x,c_{k})$ for $k=N-1,N-2,...,1$%
\ recursively as a problem of optimal irreversible switching times as follows.
Analogously to Section 2 in Azcue and Muler \cite{AM Switching}, consider the
decision-time problem with obstacle function $V^{\mathcal{G}}(x,c_{k+1})$.
Given any initial surplus $x\geq0$ and $c_{k}\in\mathcal{G}$, take the
strategy that pays dividends at constant rate $c_{k}$ up to the stopping time
$T_{k}\geq0$.
Define%
\begin{equation}
V^{\mathcal{G}}(x,c_{k})=\sup_{T_{k}}\left(  \mathbb{E}\left[  \int_{0}%
^{T_{k}\wedge\tau}e^{-qs}c_{k}ds\right]  +\mathbb{E}[e^{-q\left(  T_{k}%
\wedge\tau\right)  }V^{\mathcal{G}}(X_{T_{k}\wedge\tau},c_{k+1})]\right)
\text{.} \label{Problema obstaculo}%
\end{equation}
The value $V^{\mathcal{G}}(x,c_{k})$ can be interpreted as the expected
discounted dividend payment at rate $c_{k}$ up to the optimal stopping time
$T_{k}\wedge\tau$ plus an exit dividend payment of $V^{\mathcal{G}}%
(X_{T_{k}\wedge\tau},c_{k+1})$ at this time. With this recursive construction,
the decision time $T_{k}$ corresponds to the time at which the admissible
strategy $C=\left(  C_{t}\right)  _{t\geq0}\in\Pi_{x,c,\overline{c}%
}^{\mathcal{G}}$ changes from $c_{k}$ to some $c_{l}$ with $l>k$. We define
$T_{N}=\infty$ because $c_{N}=\overline{c}$ is the maximum possible dividend rate.

Let us define the operator%
\begin{equation}
\mathcal{L}_{c}(v)(x):=c+(p-c)v^{\prime}(x)-(q+\beta)v(x)+\beta\int
\nolimits_{0}^{x}v(x-\alpha)dF(\alpha)\text{,} \label{Lc}%
\end{equation}
then by Lemma \ref{V en cbarra} and Proposition \ref{lim l0/c},
$V^{\mathcal{G}}(\cdot,c_{N})$ is the unique solution of the
integro-differential equation $\mathcal{L}_{\overline{c}}(v)(x)=0$ in
$[0,\infty)$ with $\lim_{x\rightarrow\infty}u(x)=\overline{c}/q$. Analogously
to the proofs of Section 3 in Azcue and Muler \cite{AM Switching}, it can be
proved that $V^{\mathcal{G}}(\cdot,c_{k})$ is a viscosity solution of the
obstacle problem
\begin{equation}
\max\{\mathcal{L}_{c_{k}}(v)(x),V^{\mathcal{G}}(x,c_{k+1})-v(x)\}=0
\label{HJB de obstaculo}%
\end{equation}
in $(0,\infty)$ and that $V^{\mathcal{G}}(\cdot,c_{k})$ is the smallest
viscosity supersolution of (\ref{HJB de obstaculo}) with $\lim_{x\rightarrow
\infty}u(x)=\overline{c}/q$.

\begin{remark}\normalfont
\label{Verificacion Obstaculo} From the previous result, we conclude that if
the value function of any admissible strategy in $\Pi_{x,c_{k},\overline{c}%
}^{\mathcal{G}}$ is a viscosity supersolution of (\ref{HJB de obstaculo}),
then it is $V^{\mathcal{G}}(x,c_{k})$.
\end{remark}

If we define the closed sets%
\begin{equation}
\mathcal{D}_{k}^{\ast}:=\{x:V^{\mathcal{G}}(x,c_{k+1})-V^{\mathcal{G}}%
(x,c_{k})=0\}\text{ for }k\leq N-1, \label{Definicion Dk}%
\end{equation}
we have that $\mathcal{D}_{k}^{\ast}$ is non-empty, because otherwise
$\lim_{x\rightarrow\infty}V^{\mathcal{G}}(x,c_{k})=c_{k}/q<\overline{c}/q$.
Given any initial surplus $x\geq0$ and $c_{k}\in\mathcal{G}$, the optimal
decision time $T_{k}^{\ast}$ is the first time at which the surplus process
$X_{t}$ hits $\mathcal{D}_{k}^{\ast}.$ We define $\mathcal{D}_{N}^{\ast
}=\varnothing$ because $c_{N}=\overline{c}$ is the maximum possible dividend rate.

We have that there exist optimal strategies $C_{x,c}^{\mathcal{G}}$ $\in
\Pi_{x,c,\overline{c}}^{\mathcal{G}}$ of the (restricted) optimization problem
(\ref{Definicion VN}) for any $(x,c)\in\lbrack0,\infty)\times\lbrack
0,\overline{c}]$ and these strategies are described by the \textit{optimal
change region }%

\begin{equation}
\mathcal{D}^{\mathcal{G}}=%
{\textstyle\bigcup\nolimits_{k=1}^{N-1}}
\mathcal{D}_{k}^{\ast}\times\{c_{k}\}\subset\lbrack0,\infty)\times\mathcal{G}
\label{Optimal change region}%
\end{equation}
in the following way:

\begin{itemize}
\item Given $(x,c)\in\lbrack0,\infty)\times\lbrack0,\overline{c}]$, take
$k_{1}=\min\{k:$ with $1\leq k\leq N$,$~c_{k}\geq c$ and $x\notin
\mathcal{D}_{k}^{\ast}\},$ and pay dividends at constant rate $c_{k_{1}}$ up
to the first time $T_{k_{1}}$ that the controlled surplus process $X_{t}$ hits
$\mathcal{D}_{k_{1}}^{\ast}.$

\item Take $k_{2}=\min\{k:$ with $k_{1}<k\leq N\ $and $X_{T_{k_{1}}}%
\notin\mathcal{D}_{k}^{\ast}\},$ and pay dividends at constant rate $c_{k_{2}%
}$ up to the first time $T_{k_{2}}$ that the controlled surplus process
$X_{t}$ hits $\mathcal{D}_{k_{2}}^{\ast},$ etc. Note, that since $1\leq
k_{1}<k_{2}<\cdots\leq N,$ the number of $k_{i}^{\prime}s$ is at most $N.$
\end{itemize}

More precisely, the optimal strategy $C_{x,c}^{\mathcal{G}}=(C_{t})_{t\geq0} $
$\in\Pi_{x,c,\overline{c}}^{\mathcal{G}}$ is given by%
\[
C_{t}=\sum_{i\geq1}c_{k_{i}}I_{t\in\lbrack T_{k_{i}}\wedge\tau,T_{k_{i+1}%
}\wedge\tau)}.
\]
Note that the optimal strategies $C_{x,c}^{\mathcal{G}}$ $\in\Pi
_{x,c,\overline{c}}^{\mathcal{G}}$ are \textit{stationary} in the state space
$[0,\infty)\times\lbrack0,\overline{c}],$ in the sense that the current
dividend rate depends only on $(X_{t},C_{t^{-}})$ $\in\lbrack0,\infty
)\times\lbrack0,\overline{c}]$ (using the notation $C_{0^{-}}=c$).

\begin{remark}
\normalfont Given a finite set $\mathcal{G},$ $\mathcal{D}_{k}^{\ast}%
\ $with $k=1,...,N$ are called the {\em optimal change sets} and the
complements $\mathcal{U}_{k}^{\ast}=[0,\infty)-\mathcal{D}_{k}^{\ast}$ are
called the {\em optimal non-change sets}. In Section \ref{Seccion Ejemplos}%
, we will find examples in which the closed sets $\mathcal{D}_{k}^{\ast}$
defined in (\ref{Definicion Dk}) are of the form $[d_{k}^{\ast},\infty)$ and
satisfy $\mathcal{D}_{k+1}^{\ast}\subset\mathcal{D}_{k}^{\ast}$. In this case
$d_{k}^{\ast}$ is called the \textit{threshold} of the set $\mathcal{D}%
_{k}^{\ast}$. We will also find examples where the closed sets $\mathcal{D}%
_{k}^{\ast}$ have two connected components and $\mathcal{D}_{k+1}^{\ast
}\nsubseteq\mathcal{D}_{k}^{\ast}$.
\end{remark}

\section{Approximation with value functions of finite ratcheting strategies}

\label{sec6}

Let us now use the value functions $V^{\mathcal{G}}$ of finite ratcheting
strategies to approximate the optimal value function $V$ as the mesh size of
$\mathcal{G}$ goes to zero. For any $n\in\mathbb{N}$\textbf{,} take the
following set in $\left[  0,\overline{c}\right]  $,{\small \ }
\begin{equation}
\mathcal{G}^{n}:=\left\{  \frac{k}{2^{n}}\overline{c}\text{ }:k=0,...,2^{n}%
\right\}  . \label{Definicion Gn}%
\end{equation}
For convenience, we add $p$ to this set in the case that $\overline{c}>p$, and
use the abbreviation $V^{n}(x,c):=V^{\mathcal{G}^{n}}(x,c)$. We will prove in
this section that $\lim_{n\rightarrow\infty}V^{n}(x,c)=V(x,c)$ and we will
study the uniform convergence of this limit.

\begin{remark}
\normalfont
\label{Remark V0}The value function of the one-step ratcheting problem
considered by Albrecher et al.\ in \cite{ABB}, which increases the dividend
payment from $0$ to $\overline{c}$ only once and for all, corresponds to
$V^{0}$ with $\mathcal{G}^{0}=\left\{  0,\overline{c}\right\}  $ in our setting.
\end{remark}

Since $V^{n}\leq V^{n+1}\leq V$, we can define
\begin{equation}
\overline{V}(x,c)=\lim_{n\rightarrow\infty}V^{n}(x,c)\text{.}
\label{Definicion Vbarra}%
\end{equation}

\begin{remark}
\normalfont
\label{Lipchitz Vbarra} From Proposition \ref{Lipchitz de Vn}, we obtain
immediately similar results for $\overline{V}$:

(1) $\overline{V}(x,c)$ is non-increasing in $c$ with $\overline
{V}(x,\overline{c})=V(x,\overline{c})$, and non-decreasing in $x$ with
$\lim_{x\rightarrow\infty}$ $\overline{V}(x,c)=\overline{c}/q$.\newline(2)
There exists a constant $K_{1}>0$ such that
\[
0\leq\overline{V}(x_{2},c_{1})-\overline{V}(x_{1},c_{2})\leq K_{1}\left[
\left(  x_{2}-x_{1}\right)  +\left(  c_{2}-c_{1}\right)  \right]
\]
for all $0\leq x_{1}\leq x_{2}$ and $0\leq c_{1}\leq c_{2}\leq\min\left\{
\overline{c},p\right\}  $.\newline(3) In the case $\overline{c}>p$, there
exist constants $K_{2\text{ }}>0$ and $K_{3}>0$ such that%
\[
0\leq\overline{V}(x_{2},c_{1})-\overline{V}(x_{1},c_{2})\leq\left[
K_{2}+\frac{K_{3}}{c_{1}-p}\right]  \left(  x_{2}-x_{1}\right)  +\left[
K_{2}+\frac{K_{3}x_{2}}{\left(  c_{1}-p\right)  ^{2}}\right]  \left(
c_{2}-c_{1}\right)
\]
for all $0\leq x_{1}\leq x_{2},$ and $p<c_{1}\leq c_{2}\leq\overline{c}$.
\end{remark}

We have the following results about uniform convergence in
(\ref{Definicion Vbarra}).

\begin{proposition}
\label{Proposicion Convergencia Uniforme VN a V}In the case $\overline{c}\leq
p,$ the sequence $V^{n}$ converges uniformly to $\overline{V}.$ In the case
$\overline{c}>p,$ the sequence $V^{n}$ converges uniformly to $\overline{V}$
in $[0,\infty)\times\lbrack c_{1},\overline{c}]$ for any $c_{1}>p.$
\end{proposition}

\textit{Proof}. Let us show that $V^{n}$ converges uniformly to $\overline{V}$
in the case $\overline{c}\leq p$, the case $\overline{c}>p$ is similar. Since
$\overline{V}$ is non-decreasing in $x$ and $\lim_{x\rightarrow\infty}$
$\overline{V}(x,\overline{c})=\overline{c}/q,$ for any $\varepsilon>0$ there
exists $x_{0}$ such that
\[
0\leq\overline{c}/q-\overline{V}(x_{0},\overline{c})\leq\varepsilon/2.
\]
Take $n_{0}$ large enough such that $\overline{V}(x_{0},\overline{c}%
)-V^{n}(x_{0},\overline{c})<\varepsilon/2$ for all $n\geq n_{0}$ and so
\[
0\leq\overline{c}/q-V^{n}(x_{0},\overline{c})\leq\varepsilon.
\]
Since $\overline{V}$ and $V^{n}$ are non-decreasing in $x$ and non-increasing
in $c$, we have%
\begin{equation}
0\leq\overline{V}(x,c)-V^{n}(x,c)\leq\overline{c}/q-V^{n}(x_{0},\overline
{c})\leq\varepsilon\label{Diferencia en x grande}%
\end{equation}
for $(x,c)\in$ $[x_{0},\infty)\times\lbrack0,\overline{c}]$ and $n\geq n_{0}$.
Consider the compact set $K=[0,x_{0}]\times\lbrack0,\overline{c}]$. For any
point $(x_{1},c_{1})\in K,$ we show first that there exists $k$ large enough
and $\eta>0$ small enough such that if $\left\Vert (x,c)-(x_{1},c_{1}%
)\right\Vert <\eta$ and $n\geq$ $k$, then%
\begin{equation}
\overline{V}(x,c)-V^{n}(x,c)<\varepsilon. \label{Diferencia en Bolas}%
\end{equation}
Indeed, by pointwise convergence at $(x_{1},c_{1})$, there exists a $k$ such
that
\[
\overline{V}(x_{1},c_{1})-V^{n}(x_{1},c_{1})<\varepsilon/3~\text{for }n\geq
k.
\]
By Proposition \ref{Lipchitz de Vn} and Remark \ref{Lipchitz Vbarra}, there
exists an $\eta>0$ such that if $\left\Vert (x,c)-(x_{1},c_{1})\right\Vert
<\eta$, then
\[
\left\vert V^{n}(x,c)-V^{n}(x_{1},c_{1})\right\vert <\varepsilon/3~\text{and
}\left\vert \overline{V}(x,c)-\overline{V}(x_{1},c_{1})\right\vert
<\varepsilon/3.
\]
Therefore, we obtain (\ref{Diferencia en Bolas}). Taking a finite covering of
the compact set $K$ by balls of radius $\eta$ we conclude that there exists an
$n_{1}$ such that $\overline{V}(x,c)-V^{n}(x,c)<\varepsilon$ for any $(x,c)\in
K$ and $n\geq n_{1}$. The result follows from (\ref{Diferencia en x grande}%
).\hfill$\blacksquare$

\begin{remark}
\normalfont
In the case $\overline{c}>p,$ we cannot prove the uniform convergence in the
set $(0,\infty)\times\lbrack p,p+\varepsilon]$ with $\varepsilon>0$ because of
the lack of the Lipschitz condition of $V^{n}$, $\overline{V}$ and $V$ at the
points in the line $(0,\infty)\times\{p\}$. However, there is pointwise convergence.
\end{remark}

\begin{theorem}
The function $\overline{V}$ defined in (\ref{Definicion Vbarra}) is the
optimal value function $V$.
\end{theorem}

\textit{Proof}. Since $\overline{V}(x,c)$ is a limit of the value functions
$V^{n}(x,c)$ of admissible strategies and $\lim_{x\rightarrow\infty}$
$\overline{V}(x,\overline{c})=\overline{c}/q,$ by virtue of Theorem
\ref{verification result} it is enough to prove that $\overline{V}$ is a
viscosity supersolution of (\ref{HJB equation}) at any point $(x_{0},c_{0})$
with $x_{0}>0$ and $c_{0}\neq p.$ Since $\overline{V}$ is non-increasing in
$c$, $\overline{V}_{c}(x_{0},c_{0})\leq0$ in the viscosity sense; so it is
sufficient to show that $\mathcal{L}(\overline{V})(x_{0},c_{0})\leq0$ in the
viscosity sense. Take a test function $\varphi\ $for viscosity supersolution
of (\ref{HJB equation}) at $(x_{0},c_{0})$, i.e.\ a continuously
differentiable function $\varphi$ with
\begin{equation}
\overline{V}(x,c)\geq\varphi(x,c)\text{ and }\overline{V}(x_{0},c_{0}%
)=\varphi(x_{0},c_{0})\text{.} \label{Comparacion}%
\end{equation}
In order to prove that $\mathcal{L}(\varphi)(x_{0},c_{0})\leq0$, consider now,
for $\gamma>0$ small enough,
\[
\varphi_{\gamma}(x,c)=\varphi(x,c)-\gamma(x-x_{0})^{2}.
\]
Given $n>0,$ let us define
\[
c_{n}:=\min\{c\in\mathcal{G}^{n}:c\geq c_{0}\},
\]%
\[
a_{n}^{\gamma}:=\min\{V^{n}(x,c_{n})-\varphi_{\gamma}(x,c_{n}):x\in
\lbrack0,x_{0}+1]\},
\]%
\[
x_{n}^{\gamma}:=\arg\min\{V^{n}(x,c_{n})-\varphi_{\gamma}(x,c_{n}):x\in
\lbrack0,x_{0}+1]\},
\]
and%
\[
b_{n}^{\gamma}:=\min\{\overline{V}(x,c_{n})-V^{n}(x,c_{n}):x\in\lbrack
0,x_{0}+1]\}.
\]
We have that $c_{n}\searrow c_{0}$ and, from Proposition
\ref{Proposicion Convergencia Uniforme VN a V}, $\lim_{n\rightarrow\infty
}a_{n}^{\gamma}=0$ and $\lim_{n\rightarrow\infty}b_{n}^{\gamma}=0$. We also
have that $\lim_{n\rightarrow\infty}~x_{n}^{\gamma}=x_{0}$ because%
\[%
\begin{array}
[c]{lll}%
0 & = & V^{n}(x_{n}^{\gamma},c_{n})-\left(  \varphi_{\gamma}(x_{n}^{\gamma
},c_{n})+a_{n}^{\gamma}\right) \\
& = & \left(  V^{n}(x_{n}^{\gamma},c_{n})-\overline{V}(x_{n}^{\gamma}%
,c_{n})\right)  +\left(  \overline{V}(x_{n}^{\gamma},c_{n})-\varphi_{\gamma
}(x_{n}^{\gamma},c_{n})\right)  -a_{n}^{\gamma}\\
& \geq & -b_{n}^{\gamma}+0-a_{n}^{\gamma}+\gamma(x_{n}^{\gamma}-x_{0})^{2}%
\end{array}
\]
and then%
\[
(x_{n}^{\gamma}-x_{0})^{2}\leq\frac{a_{n}^{\gamma}+b_{n}^{\gamma}}{\gamma
}\rightarrow0~\text{as }n\rightarrow\infty.
\]

Note that $\overline{\varphi}^{n}(\cdot)=\varphi_{\gamma}(\cdot,c_{n}%
)+a_{n}^{\gamma}$ is a test function for viscosity supersolution of
$V^{n}(\cdot,c_{n})$ in equation (\ref{HJB de obstaculo}) at the point
$x_{n}^{\gamma}$ because%

\[
\varphi_{\gamma}(x_{n}^{\gamma},c_{n})+a_{n}^{\gamma}=V^{n}(x_{n}^{\gamma
},c_{n})\text{ and }\varphi_{\gamma}(x,c_{n})+a_{n}^{\gamma}\leq V^{n}%
(x,c_{n})\text{ for }x\in\lbrack0,x_{0}+1].
\]
Hence, we obtain $\mathcal{L}_{c_{n}}(\overline{\varphi}^{n})(x_{n}^{\gamma
})\leq0$. Since $(x_{n}^{\gamma},c_{n})\rightarrow(x_{0},c_{0})$,
$\overline{\varphi}^{n}(\cdot)=\varphi_{\gamma}(\cdot,c_{n})+a_{n}^{\gamma
}\rightarrow\varphi_{\gamma}(\cdot,c_{0})$ as $n\rightarrow\infty$ and
$\varphi_{\gamma}$ is continuously differentiable, one gets
\[
\mathcal{L}(\varphi_{\gamma})(x_{0},c_{0})=\lim_{n\rightarrow\infty
}\mathcal{L}_{c_{n}}(\overline{\varphi}^{n})(x_{n}^{\gamma})\leq0.
\]
Finally, as
\[
\partial_{x}\varphi_{\gamma}(x_{0},c_{0})=\partial_{x}\varphi(x_{0},c_{0})
\]
and $\varphi_{\gamma}\nearrow\varphi$ as $\gamma\searrow0$, we obtain that
$\mathcal{L}(\varphi)(x_{0},c_{0})\leq0$ and the result follows.\hfill
$\blacksquare$

\section{Numerical illustrations \label{Seccion Ejemplos}}

In this section we present some examples in which we approximate the optimal
ratcheting value $V$ by the (optimal) finite ratcheting function $\widehat
{V}:=V^{8}\ $corresponding to the finite set $\mathcal{G=G}^{8}$ with 256 possible positive values for the dividend rate defined in
(\ref{Definicion Gn}). In the examples, we will compare the value functions
$V^{0}(x,0),\ \widehat{V}(x,0)\ $and $V^{NR}(x);$ clearly, by Remarks
\ref{Optima sin ratcheting} and \ref{Remark V0}, we have that $V^{0}%
(x,0)\leq\widehat{V}(x,0)\leq V^{NR}(x)$ for all $x\geq0$. It also follows
from Proposition \ref{Lipchitz de Vn} that
\[
\lim_{x\rightarrow\infty}V^{0}(x,0)=\lim_{x\rightarrow\infty}\widehat
{V}(x,0)=\lim_{x\rightarrow\infty}V(x,0)=\lim_{x\rightarrow\infty}%
V^{NR}(x)=\overline{c}/q.
\]

Let us describe first how we obtain the optimal finite ratcheting function for
any finite set $\mathcal{G}$. Since an optimal finite ratcheting
strategy exists for any finite set $\mathcal{G}$ and it is associated to the optimal
change region $\mathcal{D}^{\mathcal{G}}$ given in
(\ref{Optimal change region}), we define, for any family of change sets
$\mathcal{D=}\left(  \mathcal{D}_{k}\right)  _{k=1,...,N-1}$ with
$\mathcal{D}_{k}\ $closed in $[0,\infty)$ the associated finite ratcheting
strategy as follows:

\begin{itemize}
\item Given $(x,c)\in\lbrack0,\infty)\times\lbrack0,\overline{c}]$, take
$k_{1}=\min\{k:$ with $1\leq k\leq N$,$~c_{k}\geq c$ and $x\notin
\mathcal{D}_{k}\},$ and pay dividends at constant rate $c_{k_{1}}$ up to the
first time $T_{k_{1}}$ that the controlled surplus process $X_{t}$ hits
$\mathcal{D}_{k_{1}}.$

\item Take $k_{2}=\min\{k:$ with $k_{1}<k\leq N\ $and $X_{T_{k_{1}}}%
\notin\mathcal{D}_{k}\}$, and pay dividends at constant rate $c_{k_{2}}$ up to
the first time $T_{k_{2}}$ that the controlled surplus process $X_{t}$ hits
$\mathcal{D}_{k_{2}},$ etc.
\end{itemize}

Denote the non-change sets as $\mathcal{U}_{k}:=[0,\infty
)\setminus \mathcal{D}_{k}$. For the value function $W^{\mathcal{D}}$ associated to the
family $\mathcal{D=}\left(  \mathcal{D}_{k}\right)_{k=1,...,N-1}$ we have the following:

\begin{itemize}
\item If $c_{N}=\overline{c},$ then $W^{\mathcal{D}}(\cdot,\overline{c})$ is
the unique solution of $\mathcal{L}_{\overline{c}}(v)=0$ with boundary
condition\\ $\lim_{x\rightarrow\infty}v(x)=\overline{c}/q.$

\item If $0\leq c_{k}<\overline{c},$ then $W^{\mathcal{D}}(\cdot
,c_{k})=W^{\mathcal{D}}(\cdot,c_{k+1})$ in $\mathcal{D}_{k}.$

\item If $p<c_{k}<\overline{c}$ and $U$ is a connected component of
$\mathcal{U}_{k}$, then $W^{\mathcal{D}}(\cdot,c_{k})$ is the unique solution
of $\mathcal{L}_{c_{k}}(v)=0$ in $U$ with boundary condition either $v(0)=0$
if $0\in U$ or $v(u_{0})=$ $W^{\mathcal{D}}(u_{0},c_{k+1})$ if $u_{0}=\inf
U\in\mathcal{D}_{k}.$

\item If $p=c_{k}\ $and $U$ is a connected component of $\mathcal{U}_{k}$, then
$W^{\mathcal{D}}(\cdot,p)$ is the unique solution of $\mathcal{L}_{p}(v)=0$ in
$U.$

\item If $0\leq c_{k}<p\ $and $U$ is a connected component of $\mathcal{U}_{k}$, then $W^{\mathcal{D}}(\cdot,c_{k})$ is the unique solution of
$\mathcal{L}_{c_{k}}(v)=0$ in $U$ with boundary condition $v(u_{0})=$
$W^{\mathcal{D}}(u_{0},c_{k+1})$ where $u_{0}=\sup U\in\mathcal{D}_{k}.$

Note that $W^{\mathcal{D}}(\cdot,c_{k})$ only depends on the sets
$D_{k},...,D_{N-1}$.
\end{itemize}

In order to find the optimal finite ratcheting strategy, we assume that the
optimal change sets $\mathcal{D}_{k}^{\ast}$ have finitely many connected
components and so $\mathcal{U}_{k}^{\ast}$ are bounded and have finitely many
connected components. We construct the optimal finite ratcheting function
$V^{\mathcal{G}}(\cdot,c_{k})$ defined in (\ref{Definicion VN}) and the
optimal change sets $\mathcal{D}_{k}^{\ast}$ for $k\leq N$ by going backward
recursively as follows: $V^{\mathcal{G}}(\cdot,c_{N})$ is the unique solution
of $\mathcal{L}_{\overline{c}}(v)=0$ with boundary condition $\lim$
$_{x\rightarrow\infty}v(x)=\overline{c}/q$. Having constructed $V^{\mathcal{G}%
}(\cdot,c_{k+1})$ and $\mathcal{D}_{k+1}^{\ast},..,\mathcal{D}_{N-1}^{\ast},$
we consider first the case in which the sets $\mathcal{D}_{k}$ and
$\mathcal{U}_{k}$ have one connected component, i.e.\ $\mathcal{D}%
_{k}=[d_{k},\infty)$, and maximize $W^{\mathcal{D}}(\cdot,c_{k})$ with the one
parameter $d_{k}$. If there exists a maximized value function $W^{\mathcal{D}%
}(\cdot,c_{k})$ and it is a viscosity supersolution of (\ref{HJB de obstaculo}%
) then, by Remark \ref{Verificacion Obstaculo}, $V^{\mathcal{G}}(\cdot
,c_{k})=W^{\mathcal{D}}(\cdot,c_{k})$ and $\mathcal{D}_{k}^{\ast}=[d_{k}%
^{\ast},\infty)$ where $d_{k}^{\ast}$ is the value for which the maximum is
attained. If this is not the case, we consider change sets $\mathcal{D}_{k}$
with two connected components and non-change sets $\mathcal{U}_{k}$ with one, 
that is $\mathcal{D}_{k}=[0,d_{k}^{1}]\cup\lbrack d_{k}^{2},\infty).$ We then
maximize $W^{\mathcal{D}}(\cdot,c_{k})$ with two parameters $d_{k}^{1}%
<d_{k}^{2},$ ; if there exists a maximized value function $W^{\mathcal{D}%
}(\cdot,c_{k})$ and it is a viscosity supersolution of (\ref{HJB de obstaculo}%
), $V^{\mathcal{G}}(\cdot,c_{k})=W^{\mathcal{D}}(\cdot,c_{k})$ and
$\mathcal{D}_{k}=[0,d_{k}^{1\ast}]\cup\lbrack d_{k}^{2\ast},\infty)$,  with 
$d_{k}^{1\ast}<d_{k}^{2\ast}$ being the values for which this maximum is attained.
If this not the case we proceed to three parameters, and so on.\\

We first consider three examples with exponentially distributed claim sizes:
two for $\overline{c}<p$ and one with $\overline{c}>p$.\newline

\begin{example}
\normalfont\label{ex1} We take $\beta=4,$ $p=2.3,$ $F(x)=1-e^{-2x},$ $q=0.1$
and $\overline{c}=1.72=\left(  3/4\right)  p$. Figure \ref{fig1} shows the
approximation $\widehat{V}(x,0)$ of the optimal value function $V(x,0)$ as a
function of $x$. Figure \ref{fig2} depicts in gray the change region
$\mathcal{D}^{\mathcal{G}^{8}}=%
{\textstyle\bigcup\nolimits_{k=1}^{2^{8}-1}}
\mathcal{D}_{k}^{\ast}\times\{\frac{k}{2^{8}}\overline{c}\}\subset
\lbrack0,\infty)\times\mathcal{G}^{8}$ (note that all the sets $\mathcal{D}%
_{k}^{\ast}$ are of the form $[d_{k}^{\ast},\infty)$ and satisfy
$\mathcal{D}_{k+1}^{\ast}\subset\mathcal{D}_{k}^{\ast}$).\newline
\begin{minipage}[c]{0.4\textwidth}
	\centering
	\includegraphics[width=2.5in]{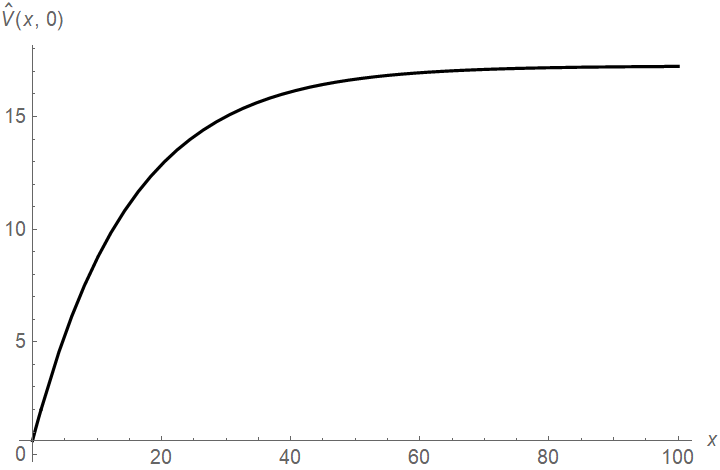}
	\captionof{figure}{{\small Value function under ratcheting as a function of initial capital $x$}}
	\label{fig1}
\end{minipage}\hspace*{0.5cm} \begin{minipage}[c]{0.4\textwidth}
	\centering
	\includegraphics[width=2.5in]{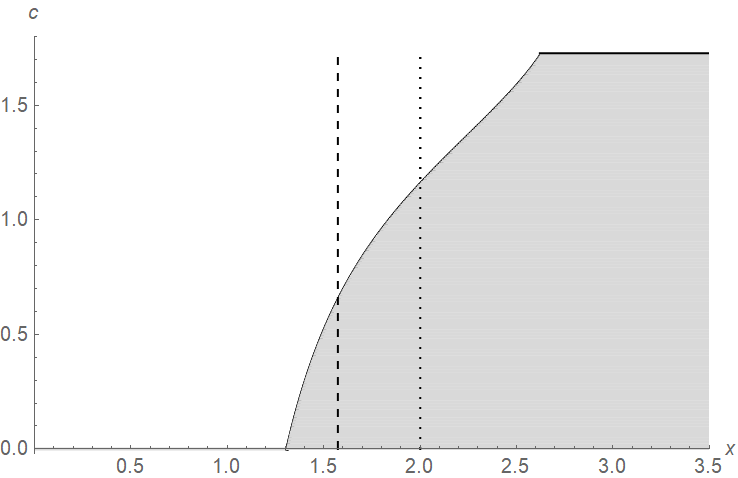}
	\captionof{figure}{{\small Change region (in gray) for the optimal ratcheting strategy}}
	\label{fig2}
\end{minipage}\\[0.2cm]


In Figure \ref{fig3}, we compare $\widehat{V}(x,0)$ with the value function
$V^{0}(x,0)$ of only one possible switch during the lifetime of the process.
We find that the optimal strategy for $V^{0}$ is given by the set
$\mathcal{D}^{\mathcal{G}^{0}}=\mathcal{D}_{1}^{\ast}\times\{0\}\subset
\lbrack0,\infty)\times\mathcal{G}^{0}$ with $\mathcal{D}_{1}^{\ast
}=[2.00,\infty)$, i.e.\ to switch to the maximal dividend rate $1.72$ as soon
as the surplus reaches the value $2$ (cf.\ the dotted line in Figure
\ref{fig2}). One sees that the difference of the value functions is
surprisingly small relative to their absolute values. That is, the improvement
from being allowed to only raise your dividend rate once rather than
ratcheting continuously is quite minor. Note also the spread of the optimal
boundary to 'enter' a dividend rate level around the one-time jump level at
$x=2$. In order to see how the number of possible switches changes the value function, Table \ref{tab1} gives the maximum error (w.r.t. the approximation $V^{8}(x,0)$) across the considered $x$-range for the respective refinements $V^{n}(x,0)$ as $n$ increases from 0 to 7. Depending on the desired accuracy one can then decide which value of $n$ is advised. In particular, $n=8$ (e.g.\ 256 possible switches) already seems to be a very satisfactory approximation of $V(x,0)$, given that the improvement over the case with 128 possible switches ($n=7$) is already quite minor.  \newline On the other hand, the efficiency loss of the expected
discounted dividend payments by ratcheting compared to the general
un-constrained dividend problem is relatively larger (yet still surprisingly
small in relative terms). Figure \ref{fig4} depicts the corresponding
difference as a function of initial capital $x$. Note that the optimal
strategy in the un-constrained problem is a threshold strategy with threshold
$x_{NR}=1.57$, i.e.\ whenever the surplus is above $x_{NR}=1.57$, dividends are
paid at the maximal rate $1.72$ and no dividends are paid when the surplus is
below that level (see e.g.\ \cite{Gerber 2006}). For illustration, $x_{NR}$ is
plotted as a dashed line in Figure \ref{fig2}. \newline

\begin{minipage}[c]{0.4\textwidth}
	\centering
	\includegraphics[width=2.5in]{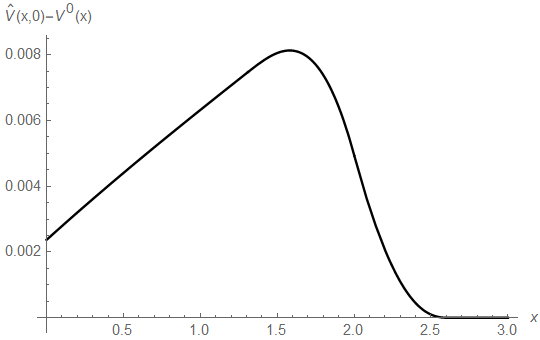}
	\captionof{figure}{{\small Improvement from switching once to general ratcheting as a function of $x$}}
	\label{fig3}
\end{minipage}\hspace*{0.5cm} \begin{minipage}[c]{0.4\textwidth}
	\centering
	\includegraphics[width=2.5in]{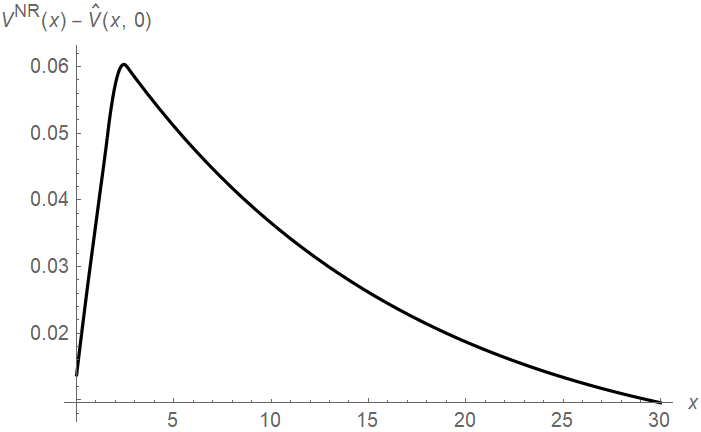}
	\captionof{figure}{{\small Improvement from the optimal ratcheting strategy to the un-constrained optimal dividend strategy as a function of $x$}}
	\label{fig4}
\end{minipage}

\begin{table}
\begin{center}
	{\footnotesize \begin{tabular}
		[c]{|c|c|c|c|c|c|c|c|c|}\hline
		$n$&${\small 0}$ & ${\small 1}$ & ${\small 2}$ & ${\small 3}$ & ${\small 4}$ &
		${\small 5}$ & ${\small 6}$ & ${\small 7}$\\\hline
		&${\small 8.05\!\cdot\!10}^{-3}$ & ${\small 2.73\!\cdot\!10}^{-3}$ & ${\small 4.96\!\cdot\!
			10}^{-4}$ & ${\small 1.17\!\cdot\!10}^{-4}$ & ${\small 2.79\!\cdot\!10}^{-4}$ &
		${\small 6.72\!\cdot\!10}^{-6}$ & ${\small 1.62\!\cdot\!10}^{-6}$ & ${\small 3.23\!\cdot\!
			10}^{-7}$\\\hline
\end{tabular}}
\end{center}\caption{$\max_x\{V^{8}(x,0)-V^{n}(x,0)\}$ for various values of $n$}
\label{tab1}
\end{table}

\end{example}

\begin{example}
\normalfont
For other parameter choices, the optimal threshold $x_{NR}$ from the
un-constrained dividend problem can in fact be lower than the first entry
point for starting dividend payments in the ratcheting case. To illustrate
this, consider $\beta=450,$ $p=100,$ $F(x)=1-e^{-5x},$ $q=0.1$ and
$\overline{c}=8$. Figure \ref{fig1a}--\ref{fig4a} then give the analogues of
Figures \ref{fig1}--\ref{fig4} of Example \ref{ex1}, and Table \ref{tab2} shows again the increasing accuracy when increasing the number of allowed switches of the dividend rate. \newline

\begin{minipage}[c]{0.4\textwidth}
	\centering
	\includegraphics[width=2.5in]{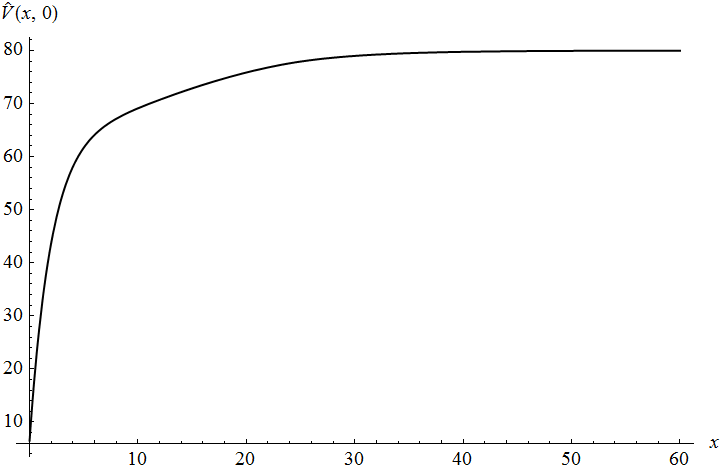}
	\captionof{figure}{{\small Value function under ratcheting as a function of initial capital $x$}}
	\label{fig1a}
\end{minipage}\hspace*{0.5cm} \begin{minipage}[c]{0.4\textwidth}
	\centering
	\includegraphics[width=2.5in]{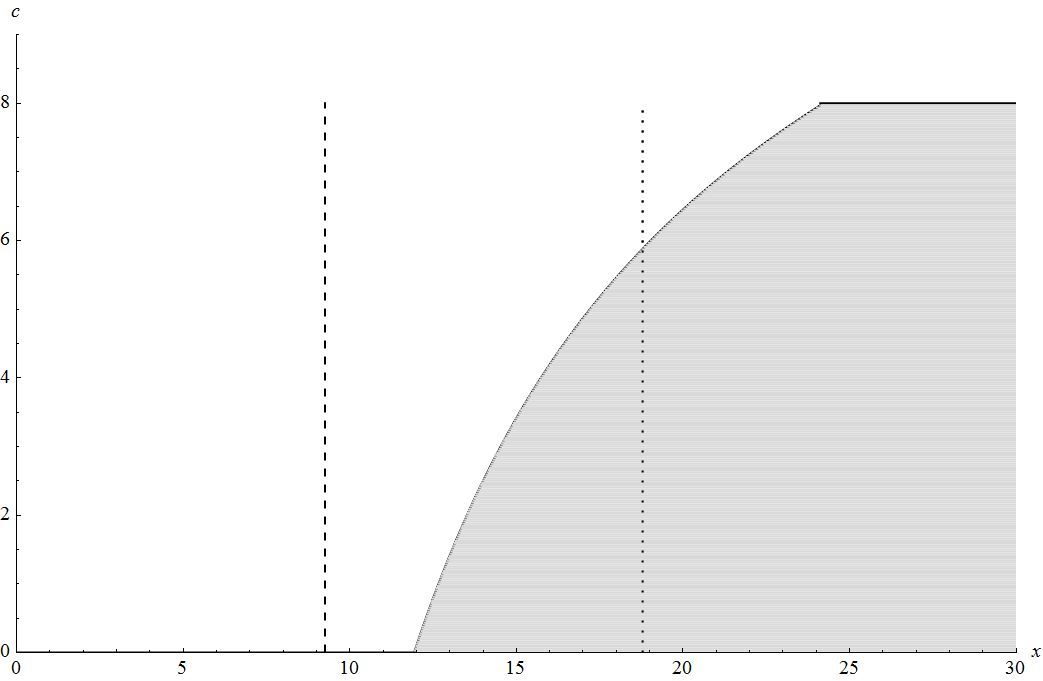}
	\captionof{figure}{{\small Change region (in gray) for the optimal ratcheting strategy}}
	\label{fig2a}
\end{minipage}

\begin{table}
	\begin{center}
		{\footnotesize \begin{tabular}
				[c]{|c|c|c|c|c|c|c|c|c|}\hline
				$n$&${\small 0}$ & ${\small 1}$ & ${\small 2}$ & ${\small 3}$ & ${\small 4}$ &
				${\small 5}$ & ${\small 6}$ & ${\small 7}$\\\hline
				&${\small 7.80\!\cdot\!10}^{-1}$ & ${\small 2.82\!\cdot\!10}^{-1}$ & ${\small 7.99\!\cdot\!
					10}^{-2}$ & ${\small 2.06\!\cdot\!10}^{-2}$ & ${\small 5.19\!\cdot\!10}^{-3}$ &
				${\small 1.28\!\cdot\!10}^{-3}$ & ${\small 3.06\!\cdot\!10}^{-4}$ & ${\small 6.11\!\cdot\!
					10}^{-5}$\\\hline
		\end{tabular}}
	\end{center}\caption{$\max_x\{V^{8}(x,0)-V^{n}(x,0)\}$ for various values of $n$}
	\label{tab2}
\end{table}

Here, the optimal strategy for $V^{0}$ is given by the set $\mathcal{D}%
^{\mathcal{G}^{0}}=\mathcal{D}_{1}^{\ast}\times\{0\}\subset\lbrack
0,\infty)\times\mathcal{G}^{0}$ with $\mathcal{D}_{1}^{\ast}=[18.79,\infty)$
and the optimal strategy in the non-ratcheting problem is a threshold strategy
with threshold $x_{NR}=9.26$, the latter being to the left of the gray region
in Figure \ref{fig2a}.\newline

\begin{minipage}[c]{0.4\textwidth}
	\centering
	\includegraphics[width=2.1in]{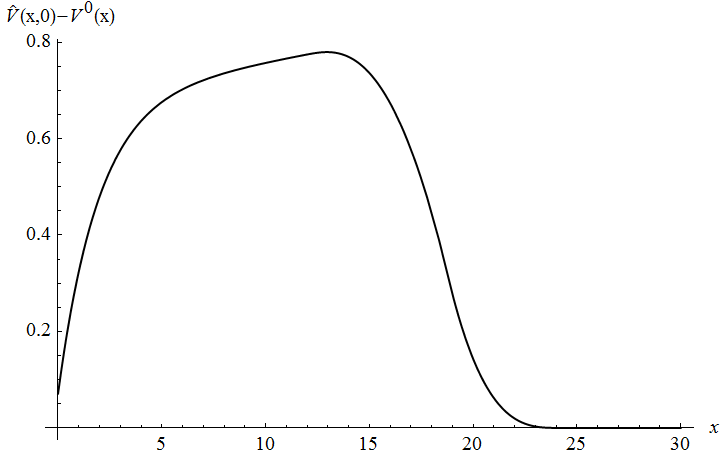}
	\captionof{figure}{{\small Improvement from switching once to general ratcheting as a function of $x$}}
	\label{fig3a}
\end{minipage}\hspace*{0.2cm} \begin{minipage}[c]{0.4\textwidth}
	\centering
	\includegraphics[width=2.1in]{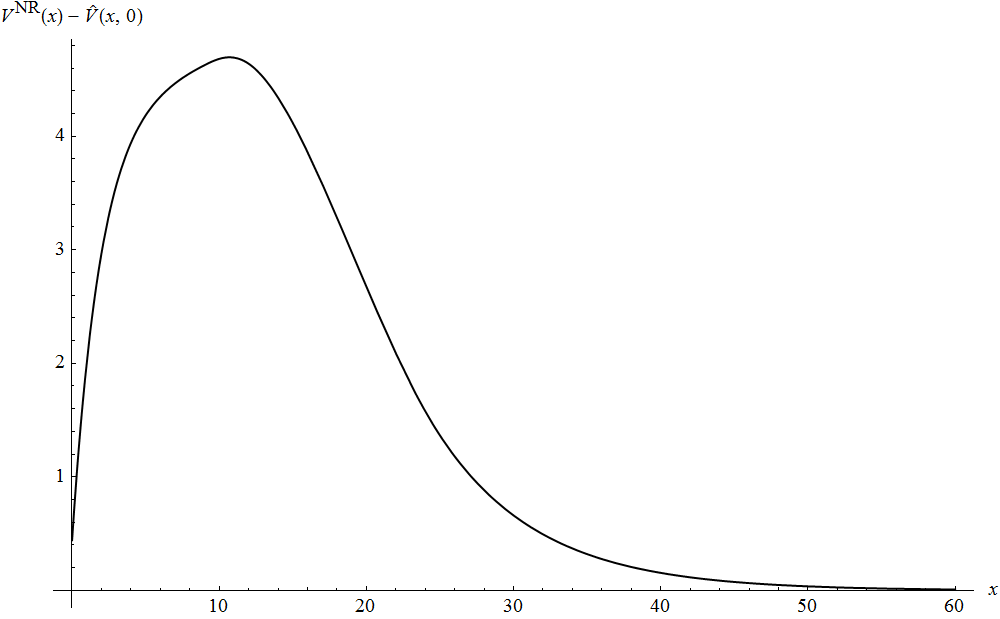}
	\captionof{figure}{{\small Improvement from the optimal ratcheting strategy to the un-constrained optimal dividend strategy as a function of $x$}}
	\label{fig4a}
\end{minipage}
\newline
\end{example}

\begin{example}
\normalfont The geometry of the change region can become more complex, if the
maximal dividend rate is allowed to exceed the premium rate $p$. As an
illustration, consider $\beta=4,$ $p=2.3,$ $F(x)=1-e^{-2x},$ $q=0.1$ and
$\overline{c}=4.6=2p$. Figure \ref{fig6} gives the approximation $\widehat
{V}(x,0)$ of the optimal value function $V(x,0)$ in this case, and the change
region $\mathcal{D}^{\mathcal{G}^{8}}$ is plotted (in gray) in Figure
\ref{fig7} (again, all the sets $\mathcal{D}_{k}^{\ast}$ are of the form
$[d_{k}^{\ast},\infty)$ and satisfy $\mathcal{D}_{k+1}^{\ast}\subset
\mathcal{D}_{k}^{\ast}$).\newline

\begin{minipage}[c]{0.4\textwidth}
	\centering
	\includegraphics[width=2.5in]{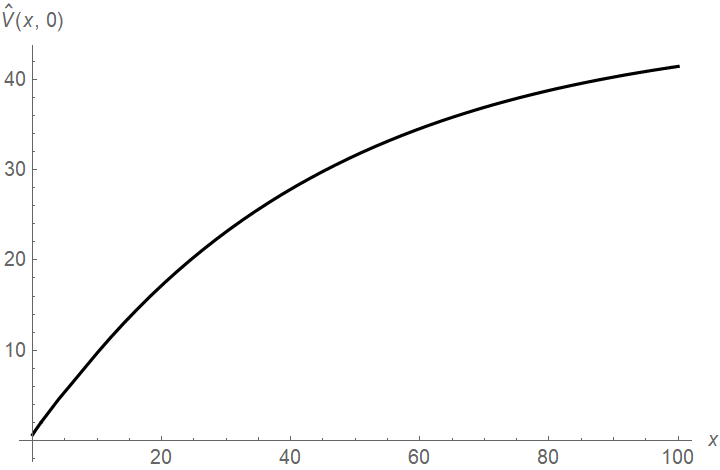}
	\captionof{figure}{{\small Value function under ratcheting as a function of initial capital $x$}}
	\label{fig6}
\end{minipage}\hspace*{0.5cm} \begin{minipage}[c]{0.4\textwidth}
	\centering
	\includegraphics[width=2.5in]{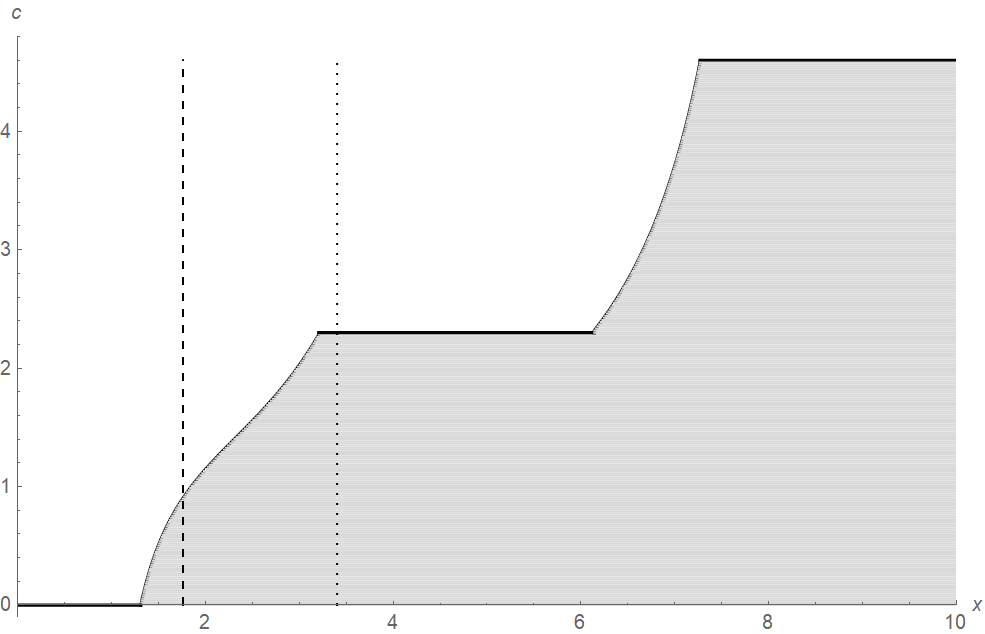}
	\captionof{figure}{{\small Change region (in gray) for the optimal ratcheting strategy}}
	\label{fig7}
\end{minipage}\newline\begin{minipage}[c]{0.4\textwidth}
	\centering
	\includegraphics[width=2.1in]{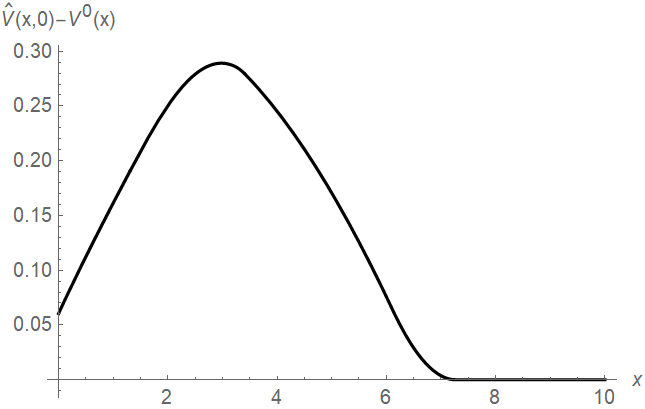}
	\captionof{figure}{{\small Improvement from switching once to general ratcheting as a function of $x$}}
	\label{fig8}
\end{minipage}\hspace*{0.2cm} \begin{minipage}[c]{0.4\textwidth}
	\centering
	\includegraphics[width=2.1in]{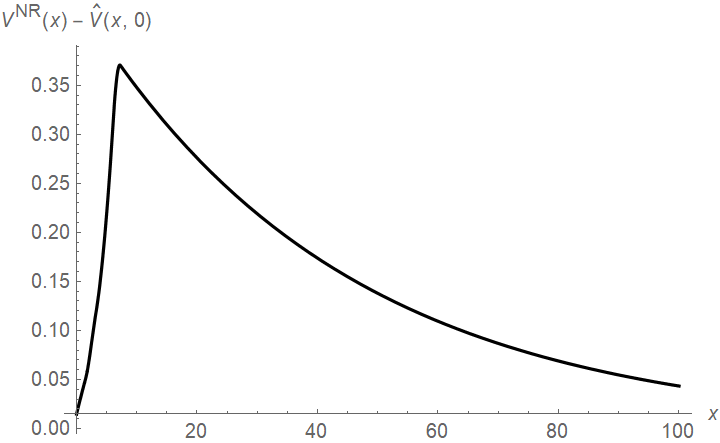}
	\captionof{figure}{{\small Improvement from the optimal ratcheting strategy to the un-constrained optimal dividend strategy as a function of $x$}}
	\label{fig9}
\end{minipage}

\begin{table}
	\begin{center}
		{\footnotesize \begin{tabular}
				[c]{|c|c|c|c|c|c|c|c|c|}\hline
				$n$&${\small 0}$ & ${\small 1}$ & ${\small 2}$ & ${\small 3}$ & ${\small 4}$ &
				${\small 5}$ & ${\small 6}$ & ${\small 7}$\\\hline
				&${\small 2.89\!\cdot\!10}^{-1}$ & ${\small 2.22\!\cdot\!10}^{-2}$ & ${\small 7.51\!\cdot\!
					10}^{-3}$ & ${\small 1.18\!\cdot\!10}^{-3}$ & ${\small 7.61\!\cdot\!10}^{-4}$ &
				${\small 5.38\!\cdot\!10}^{-5}$ & ${\small 1.44\!\cdot\!10}^{-5}$ & ${\small 2.55\!\cdot\!
					10}^{-6}$\\\hline
		\end{tabular}}
	\end{center}\caption{$\max_x\{V^{8}(x,0)-V^{n}(x,0)\}$ for various values of $n$}
	\label{tab3}
\end{table}

\bigskip

One observes that now we have a region of the free boundary where the dividend
rate $c$ is not increased for growing $x$. Concretely, it takes a much larger
surplus value until it is optimal to start making use of the possibility to
pay out dividends at a higher rate than the one of incoming premiums. Figure
\ref{fig8} again shows the efficiency gain from the optimal one-switch
strategy (which in this case is characterized by the set $\mathcal{D}%
^{\mathcal{G}^{0}}=\mathcal{D}_{1}^{\ast}\times\{0\}\subset\lbrack
0,\infty)\times\mathcal{G}^{0}$ with $\mathcal{D}_{1}^{\ast}=[3.40,\infty)$)
to the general ratcheting strategy, and Table \ref{tab3} shows again the quality improvement of the approximation when allowing more and more switches. Figure \ref{fig9}, on the other hand,
depicts the difference of the value functions of the optimal dividend problem
without and with ratcheting. In this case, the optimal threshold for the
non-constrained case is $x_{NR}=1.76;$ since $\overline{c}>p,$ this means that
the optimal strategy is to pay no dividends if the current surplus is less
than $x_{NR}$, to pay dividends at maximum possible rate $\overline{c}$ if the
current surplus is greater than $x_{NR}$ and to pay dividends at rate $p$ if
the current surplus coincides with $x_{NR}$.\newline

\end{example}

\begin{remark}
\normalfont
The numerical approximation suggests that, in the examples above, the optimal
ratcheting strategy is given by the optimal free boundary
\[
G=\{(x,C^{\ast}(x)):x\geq0\},
\]
which separates the non-change region with the change region; here $C^{\ast}$
is a non-decreasing function with $C^{\ast}(x)=0$ for $x$ small and $C^{\ast
}(x)=\overline{c}$ for $x$ large enough. More precisely,

\begin{itemize}
\item If the initial values are $(x,c)$ with $c<C^{\ast}(x)$, the optimal
strategy is to pay dividends at rate $C^{\ast}(x)$.

\item If the initial values are $(x,c)$ with $c>C^{\ast}(x),$ the optimal
strategy is to pay dividends at rate $c$ until the controlled trajectory
$(X_{t}^{C}-ct,c)$ in the state space reaches the free boundary $G$.

\item If the initial values $(x,c)$ are on the free boundary $G$ with $c<p$,
the optimal strategy is to pay dividends at rate $C^{\ast}(x)$. In this case
the trajectory in the state space $\left(  X_{t}^{C},C^{\ast}(X_{t}%
^{C})\right)  \ $remains on the free boundary $G$ until either ruin occurs or
a next claim arrives.

\item If the initial values $(x,c)$ are on the free boundary $G$ with $c=p$,
the optimal strategy is to pay dividends at rate $p$. In this case the
trajectory in the state space $\left(  X_{s}^{C},C^{\ast}(X_{s}^{C})\right)
\ $is constant $(x,p)$ until either ruin occurs or a next claim arrives.

\item If the initial values $(x,c)$ are on the free boundary $G$ with $c>p$,
the optimal strategy is to pay dividends at rate $c$. In this case the
trajectory in the state space falls immediately into the non-change region.
\end{itemize}
\end{remark}

\medskip

\begin{example}
\label{ex4}\normalfont As a final example, we would like to see how the
ratcheting constraint changes the un-constrained dividend problem in the case
where a band strategy is optimal for the latter. To that end, choose
$\beta=10,$ $p=21.4,$ $F(x)=1-e^{-x}(1+x),$ $q=0.1$ and $\overline
{c}=16.5=\left(  3/4\right)  p$ (cf.\ \cite[Sec.10.1]{AM05}). Figure
\ref{fig11} depicts the approximation $\widehat{V}(x,0)$ of the optimal value
function $V(x,0)$ for this choice of parameters, while Figure \ref{fig12}
shows in gray the change region $\mathcal{D}^{\mathcal{G}^{8}}$, and Figures
\ref{fig14} and \ref{fig15} compare $\widehat{V}(x,0)$ again with the
one-switch value function $V^{0}(x,0)$ and the un-constrained value function
$V^{NR}(x,0)$, respectively. Table \ref{tab4} depicts the approximation improvement when increasing the number of possible switches.\newline From Figure \ref{fig12} one sees that in
this case the sets $\mathcal{D}_{k}^{\ast}=[0,d_{k}^{1\ast}]\cup\lbrack
d_{k}^{2\ast},\infty)$ have two components with $d_{k}^{1\ast}<d_{k+1}^{1\ast
}$ and $d_{k}^{2\ast}<d_{k+1}^{2\ast}$ and so $\mathcal{D}_{k+1}^{\ast
}\nsubseteqq\mathcal{D}_{k}^{\ast}$. In Figure \ref{fig13}, we show the
derivative of $\widehat{V}(x,0)$ for small values of $x$. Note that the
function $\widehat{V}(x,0)$ is continuous but not differentiable at the point
$d_{0}^{1\ast}=0.076$ because the optimal strategy consists of paying
dividends at maximum rate $\overline{c}$ if the current surplus is less than
or equal to $d_{0}^{1\ast}$, but paying no dividends if the current surplus is
slightly above $d_{0}^{1\ast}$. Only for much larger values of the initial capital
$x$ it is again optimal to pay dividends. \newline

\begin{minipage}[c]{0.4\textwidth}
	\centering
	\includegraphics[width=2.1in]{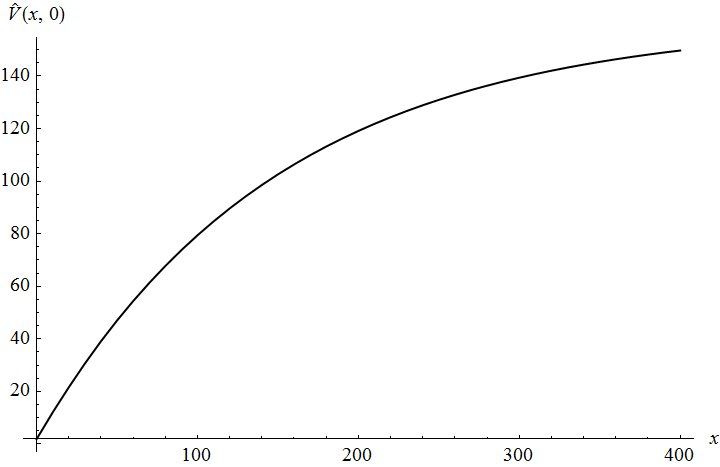}
	\captionof{figure}{{\small Value function under ratcheting as a function of initial capital $x$}}
	\label{fig11}
\end{minipage}\hspace*{0.3cm} \begin{minipage}[c]{0.4\textwidth}
	\centering
	\includegraphics[width=2.1in]{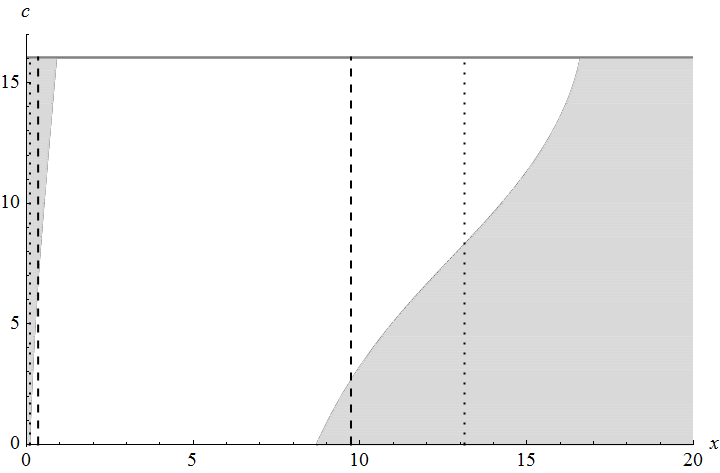}
	\captionof{figure}{{\small Change region (in gray) for the optimal ratcheting strategy}}
	\label{fig12}
\end{minipage}\newline\begin{minipage}[c]{0.4\textwidth}
	\centering
	\includegraphics[width=2.1in]{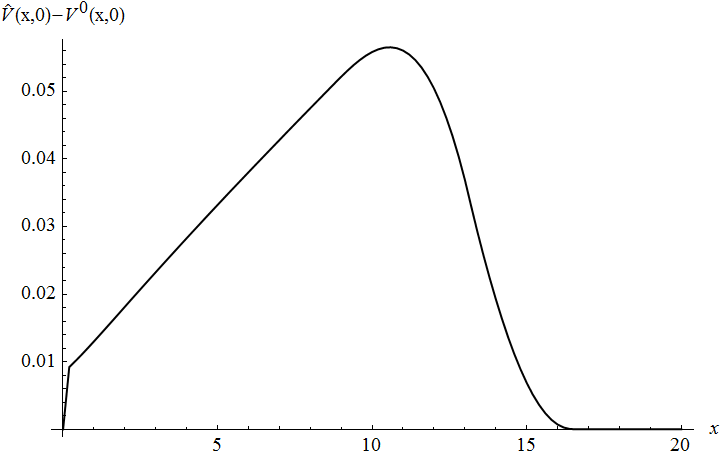}
	\captionof{figure}{{\small Improvement from switching once to general ratcheting as a function of $x$}}
	\label{fig14}
\end{minipage}\hspace*{0.2cm} \begin{minipage}[c]{0.4\textwidth}
	\centering
	\includegraphics[width=2.1in]{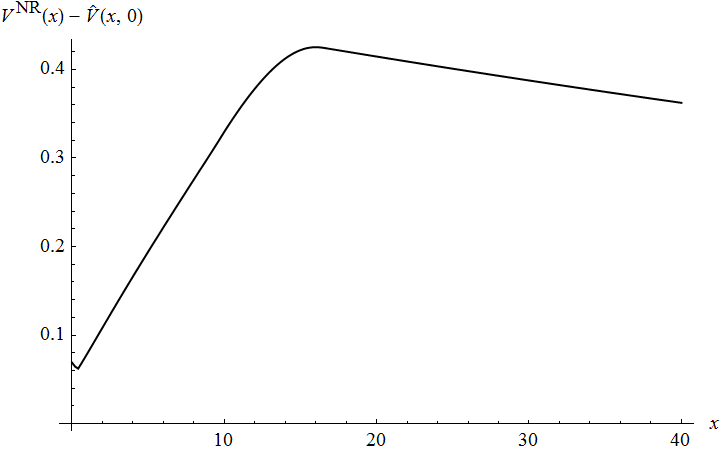}
	\captionof{figure}{{\small Improvement from the optimal ratcheting strategy to the un-constrained optimal dividend strategy as a function of $x$}}
	\label{fig15}
\end{minipage}

\begin{center}
	\begin{minipage}[c]{0.5\textwidth}
	\centering
	\includegraphics[width=1.8in]{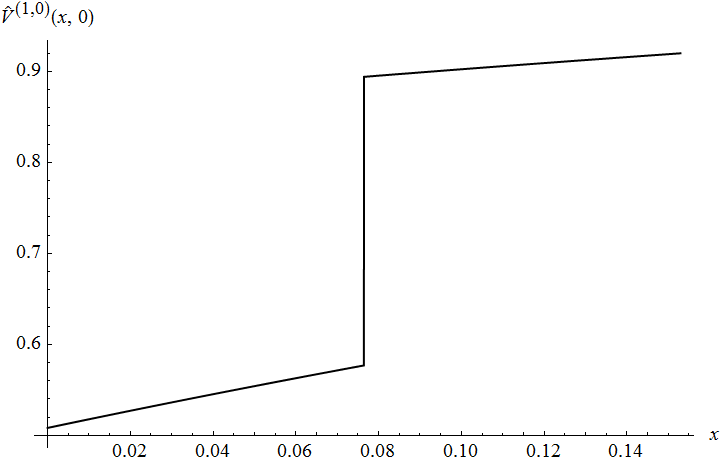}
	\captionof{figure}{{\small The derivative of $\widehat{V}(x,0)$ with respect to initial capital $x$}}
	\label{fig13}
\end{minipage}
\end{center}

\begin{table}
	\begin{center}
		{\footnotesize \begin{tabular}
				[c]{|c|c|c|c|c|c|c|c|c|}\hline
				$n$&${\small 0}$ & ${\small 1}$ & ${\small 2}$ & ${\small 3}$ & ${\small 4}$ &
				${\small 5}$ & ${\small 6}$ & ${\small 7}$\\\hline
			&	${\small 5.65\!\cdot\!10}^{-2}$ & ${\small 9.45\!\cdot\!10}^{-3}$ & ${\small 2.45\!\cdot\!
					10}^{-3}$ & ${\small 5.97\!\cdot\!10}^{-4}$ & ${\small 1.47\!\cdot\!10}^{-4}$ &
				${\small 3.63\!\cdot\!10}^{-5}$ & ${\small 8.67\!\cdot\!10}^{-6}$ & ${\small 1.72\!\cdot\!
					10}^{-6}$\\\hline
		\end{tabular}}
	\end{center}\caption{$\max_x\{V^{8}(x,0)-V^{n}(x,0)\}$ for various values of $n$}
	\label{tab4}
\end{table}

This is exactly the extension of the band strategy of the un-constrained case
(which consists of paying dividends at maximum possible rate $\overline{c}$ if
the current surplus is in the set $[0,0.22]\cup\lbrack12.05,\infty)$, cf.\ the
dashed lines in Figure \ref{fig12}, and paying no dividends elsewhere -- so
not a threshold strategy anymore) to the ratcheting case. The optimal strategy
for $V^{0}$ (i.e.\ when only one switch of dividend rates is possible) is
given by the set $\mathcal{D}^{\mathcal{G}^{0}}=\mathcal{D}_{1}^{\ast}%
\times\{0\}\subset\lbrack0,\infty)\times\mathcal{G}^{0}$, where $\mathcal{D}%
_{1}^{\ast}=[0,0.10]\cup\lbrack13.13,\infty)$ now has two connected components
(cf.\ dotted lines in Figure \ref{fig12}).

\end{example}

\begin{remark}
\normalfont
In Example \ref{ex4} the optimal ratcheting strategy is given by two optimal
free boundaries
\[
G_{1}=\{(x,C_{1}^{\ast}(x)):x\geq0\}\text{ and }G_{2}=\{(x,C_{2}^{\ast
}(x)):x\geq0\}
\]
which bound the non-change region on the left and on the right respectively;
here $C_{1}^{\ast}$ and $C_{2}^{\ast}$ are non-decreasing functions with
$C_{2}^{\ast}$ $\leq C_{1}^{\ast}$ and $C_{i}^{\ast}(x)=0$ for $x$ small and
$C_{i}^{\ast}(x)=\overline{c}$ for $x$ large enough for $i=1,2$. More precisely,

\begin{itemize}
\item If the initial values are $(x,c)$ with $c\geq C_{1}^{\ast}(x)$, the
optimal strategy is to pay dividends at maximum rate $\overline{c}$.

\item If the initial values are $(x,c)$ with $c<C_{2}^{\ast}(x)$, the optimal
strategy is to pay dividends at rate $C_{2}^{\ast}(x)$.

\item If the initial values $(x,c)$ are on the free boundary $G_{2}$ but not
in $G_{1}$, the optimal strategy is to pay dividends at rate $C_{2}^{\ast}%
(x)$; in this case the trajectory in the state space $\left(  X_{t}^{C}%
,C_{2}^{\ast}(X_{t}^{C})\right)  \ $remains on the free boundary $G_{2}$ until
either ruin occurs or the next claim arrives.

\item If the initial values $(x,c)$ are in the non-change region
(i.e.\ $C_{2}^{\ast}(x)<c<C_{1}^{\ast}(x)$), then the optimal strategy is to
pay dividends at rate $c$ until the controlled trajectory $(X_{t}^{C}-ct,c)$
in the state space exits the non-change region.
\end{itemize}
\end{remark}

\section{Conclusion}

\label{seccon} In this paper we solved the general problem of identifying
optimal dividend strategies in an insurance risk model under the additional
constraint that the dividend rate needs to be non-decreasing over time. We
showed that the value function is the unique viscosity solution of a
two-dimensional Hamilton-Jacobi-Bellman equation and can be approximated
arbitrarily closely by optimal strategies for finitely many possible dividend
rates. The analysis is considerably more complicated when the maximal dividend
rate is allowed to exceed the incoming premium rate. We derived the free
boundaries and optimal strategies numerically for a number of concrete cases
with exponential and Gamma claim sizes, and the results illustrate that the
value function with ratcheting is not much lower than the one without the
ratcheting constraint. Also, a comparison shows that the previously studied
one-switch strategy performs remarkably well, i.e.\ the further improvement in
the case of general ratcheting is typically not substantial. We also showed
that for parameter settings where a band strategy is optimal in the
non-constrained case, the band-type structure remains optimal for the
ratcheting solution, then with two free boundaries in the domain of initial
surplus and initial dividend rate.

\section{Appendix \ }

\subsection{Proof of Proposition \ref{Proposition Global Lipschitz zone}}

Proposition \ref{Proposition Global Lipschitz zone} follows from the two
following Lemmas.

\begin{lemma}
\label{Lemma Liptchitz x, c menor a p}There exists $K_{1}>0$ such that%
\[
V(x_{2},c)-V(x_{1},c)\leq K_{1}\left(  x_{2}-x_{1}\right)
\]

for all $0\leq x_{1}\leq x_{2}$ and $c\leq\min\left\{  \overline{c},p\right\}
.$
\end{lemma}

\emph{Proof.} Take $\varepsilon>0$ and $C\in\Pi_{x_{2},c,\overline{c}}$ such
that%
\begin{equation}
J(x_{2};C)\geq V(x_{2},c)-\varepsilon. \label{casi optima x2}%
\end{equation}
Then the associated control process is given by%
\[
X_{t}^{C}=x_{2}+\int_{0}^{t}(p-C_{s})ds-\sum_{i=1}^{N_{t}}U_{i}.
\]
Let $\tau$ be the ruin time of the process $X_{t}^{C}$. \bigskip Assume first
that $\overline{c}\leq p$ and define $\widetilde{C}\in\Pi_{x_{1}%
,c,\overline{c}}$ as $\widetilde{C}_{t}=C_{t}$, where
\[
\text{ }X_{t}^{\widetilde{C}}=x_{1}+\int_{0}^{t}(p-C_{s})ds-\sum_{i=1}^{N_{t}%
}U_{i}.
\]
For the ruin time $\widetilde{\tau}\leq\tau$ of the process $X_{t}%
^{\widetilde{C}}$, it holds $X_{t}^{C}-X_{t}^{\widetilde{C}}=x_{2}-x_{1}$ for
$t\leq\widetilde{\tau}$. Since $\overline{c}\leq p$, ruin can occur only at
the arrival of a claim. Hence, using (\ref{casi optima x2}) we have
\begin{equation}%
\begin{array}
[c]{lll}%
V(x_{2},c)-V(x_{1},c) & \leq & J(x_{2};C)-J(x_{1};\widetilde{C})+\varepsilon\\
& = & \mathbb{E}[\int_{\widetilde{\tau}}^{\tau}C_{s}e^{-qs}ds]+\varepsilon\\
& \leq & \mathbb{E}[\sum_{j=1}^{\infty}I_{\left\{  \widetilde{\tau}=\tau
_{j}\text{ and }\tau>\tau_{j}\right\}  }\left(  \int_{\tau_{j}}^{\tau}%
C_{s}e^{-qs}ds\right)  ]+\varepsilon\\
& \leq & \frac{\overline{c}}{q}\,\mathbb{E}[\sum_{j=1}^{\infty}e^{-q\tau_{j}%
}I_{\left\{  \widetilde{\tau}=\tau_{j}\text{ and }\tau>\tau_{j}\right\}
}]+\varepsilon.
\end{array}
\label{diferencia de vs}%
\end{equation}

With the definitions
\begin{equation}
\mathcal{U}_{j-1}:=\sum_{i=1}^{j-1}U_{i}~\text{and~}A_{t}^{C}:=\int_{0}%
^{t}(p-C_{s})ds, \label{U y At}%
\end{equation}
we have
\[
\left\{  \widetilde{\tau}=\tau_{j}\text{ and }\tau>\tau_{j}\right\}  =\left\{
x_{2}+A_{\tau_{j}}^{C}-\mathcal{U}_{j-1}\geq U_{j}~>x_{1}+A_{\tau_{j}}%
^{C}-\mathcal{U}_{j-1}\right\}  ,
\]
and by the i.i.d.\ assumptions $\tau_{j}$, $U_{j}$ and $\mathcal{U}_{j-1}$ are
mutually independent. This implies%

\begin{equation}%
\begin{array}
[c]{l}%
\mathbb{E}[\sum_{j=1}^{\infty}e^{-q\tau_{j}}I_{\left\{  \widetilde{\tau}%
=\tau_{j}\text{,}\tau>\tau_{j}\right\}  }]\\%
\begin{array}
[c]{ll}%
\leq & K(x_{2}-x_{1})\beta\sum_{j=1}^{\infty}\left[  \int_{0}^{\infty}%
e^{-qt}\left(  \frac{\beta^{j-1}t^{j-1}}{(j-1)!}\right)  e^{-\beta t}dt\right]
\\
\leq & K\frac{\beta}{q}(x_{2}-x_{1}),
\end{array}
\end{array}
\label{Acotacion_EsperanzaDifx:RuinaSalto}%
\end{equation}
because $F(A_{t}+x_{2}-\mathcal{U}_{j-1})-F(x_{1}+A_{t}-\mathcal{U}_{j-1})\leq
K(x_{2}-x_{1})$. From (\ref{diferencia de vs}) and
(\ref{Acotacion_EsperanzaDifx:RuinaSalto}) we get the result with
$K_{1}=K\beta\overline{c}/q^{2}$.

\bigskip

Consider now $c\leq p<\overline{c}.$ The main difference in this case is that
ruin can occur not only at the arrival of a claim but also if dividends are
paid with current surplus zero at a rate greater than $p$.

Let us prove first the result for $c=p$. Consider $C\in\Pi_{x_{2}%
,p,\overline{c}}$ as in (\ref{casi optima x2}) and
\begin{equation}
T=\min\left\{  t:\int_{0}^{t}\left(  C_{s}-p\right)  ds=x_{2}-x_{1}\right\}  .
\label{Definicion T_crucetrayectorias}%
\end{equation}
We put $T=\infty$ in the event
\[
\int_{0}^{\tau}\left(  C_{s}-p\right)  ds<x_{2}-x_{1}.
\]
Define $\overline{C}\in\Pi_{x_{1},p,\overline{c}}$ as follows: $\overline
{C}_{t}=p$ for $t\leq T$ and then $\overline{C}_{t}=C_{t}$ and $\overline
{\tau}\leq\tau$ as the ruin time of the controlled process $X_{t}%
^{\overline{C}}$. Note that if $T\leq\overline{\tau}$ we have $X_{T}^{C}%
=X_{T}^{\overline{C}}$ because%

\[
X_{T}^{C}-X_{T}^{\overline{C}}=x_{2}-x_{1}+\int_{0}^{T}(p-C_{s})ds=0
\]
and so $X_{t}^{C}=X_{t}^{\overline{C}}$ for $T\leq t\leq\overline{\tau}=\tau$.
In the event that $T>\overline{\tau}$, we have $0<X_{t}^{C}-X_{t}%
^{\overline{C}}\leq x_{2}-x_{1}$ for all $t\leq\overline{\tau}$; also
$\overline{\tau}$ coincides with the arrival of a claim since $\overline
{C}_{s}=p$ for $s\leq\overline{\tau}$. Therefore, from
(\ref{Definicion T_crucetrayectorias}) and using the proof of
(\ref{Acotacion_EsperanzaDifx:RuinaSalto}) we can write%

\begin{equation}%
\begin{array}
[c]{l}%
V(x_{2},p)-V(x_{1},p)\\%
\begin{array}
[c]{ll}%
\leq & J(x_{2};C)-J(x_{1};\overline{C})+\varepsilon\\
= & \mathbb{E}[I_{T\leq\overline{\tau}}\left(  \int_{0}^{T}\left(
C_{s}-p\right)  e^{-qs}ds\right)  ]\\
& +\mathbb{E}\left[  I_{T>\overline{\tau}}\int_{0}^{\overline{\tau}}\left(
C_{s}-\overline{C}_{s}\right)  e^{-qs}ds\right]  +\mathbb{E}\left[
I_{T>\overline{\tau}}\int_{\overline{\tau}}^{\tau}C_{s}e^{-qs}ds\right]
+\varepsilon\\
\leq & 2(x_{2}-x_{1})+\mathbb{E}[I_{\overline{\tau}\leq T}\sum_{j=1}^{\infty
}I_{\left\{  \overline{\tau}=\tau_{j},\tau>\tau_{j}\right\}  }\left(
\int_{\tau_{j}}^{\tau}C_{s}e^{-qs}ds\right)  ]+\varepsilon\\
\leq & \left(  2+\overline{c}~K\frac{\beta}{q^{2}}\right)  (x_{2}%
-x_{1})+\varepsilon
\end{array}
\end{array}
\label{cota lipschitz en p_cbarra mayor p}%
\end{equation}
and so we get the result with $K_{1}=2+\overline{c}~K\beta/q^{2}$.

Let us consider now the case $c<p<\overline{c}$ , $C\in\Pi_{x_{2}%
,c,\overline{c}}$ as in (\ref{casi optima x2}) and define
\[
T_{1}=\min\left\{  t:C_{t}\geq p\right\}  ;
\]
if $C_{t}\leq p$ for all $t\leq\tau$ then $T_{1}=\infty$.

Since $V(\cdot,p)$ is non-decreasing and continuous, we can find (as in Lemma
1.2 of \cite{AM Libro}) an increasing sequence $(y_{i})\ $with $y_{1}=0$ such
that if $y\in\lbrack y_{i},y_{i+1})$ then $0\leq V(y,p)-V(y_{i},p)\leq
\varepsilon/2;$ consider admissible strategies $\widehat{C}^{i}\in\Pi
_{y_{i},p,\overline{c}}$ such that $V(y_{i},p)-J(y_{i},\widehat{C}^{i}%
)\leq\varepsilon/2$. Let us define the dividend payment strategy $\overline
{C}\in\Pi_{x_{1},c,\overline{c}}$ as follows: $\overline{C}_{t}=C_{t}$ for
$t<T_{1}$ and $\overline{C}_{t}=\widehat{C}_{t-T_{1}}^{i}$ for $t\geq T_{1}$
in the case that $X_{T_{1}}^{C}\in\lbrack y_{i},y_{i+1})$; note that, with
this definition, the strategy $\overline{C}$ turns out to be Borel measurable
and so it is admissible. With arguments similar to the ones used before, we
obtain%
\[
V(x_{2},p)-V(x_{1},p)\leq\left(  2+2\overline{c}~K\frac{\beta}{q^{2}}\right)
(x_{2}-x_{1})\text{. }\blacksquare
\]

\begin{lemma}
There exists $K_{2}>0$ such that%
\[
0\leq V(x,c_{1})-V(x,c_{2})\leq K_{2}\left(  c_{2}-c_{1}\right)
\]
for all $x\geq0$ and $0\leq c_{1}\leq c_{2}\leq\min\left\{  \overline
{c},p\right\}  .$
\end{lemma}

\emph{Proof.} Take $\varepsilon>0$ and $C\in\Pi_{x,c_{1},\overline{c}}$ such that%

\begin{equation}
J(x;C)\geq V(x,c_{1})-\varepsilon\label{casiOptimac1}%
\end{equation}
and define the stopping time
\begin{equation}
\widehat{T}=\min\{t:C_{t}\geq c_{2}\}. \label{Definicion TauSombrero}%
\end{equation}
Recall that $\tau$ is the ruin time of the process $X_{t}^{C}$. Consider first
the case $\overline{c}\leq p$ and define $\widetilde{C}\in\Pi_{x,c_{2}%
,\overline{c}}$ as $\widetilde{C}_{t}=c_{2}I_{t<\widehat{T}}+C_{t}%
I_{t\geq\widehat{T}}$; denote by $X_{t}^{C}$ the associated controlled surplus
process and by $\overline{\tau}\leq\tau$ the corresponding ruin time. Since
$\overline{c}\leq p$, both $X_{t}^{C}$ and $X_{t}^{\widetilde{C}}$ are
non-decreasing between claim arrivals, and ruin can only occur at the arrival
of a claim. We also have that $\widetilde{C}_{s}-C_{s}\leq c_{2}-c_{1}$. We
can write%
\begin{equation}%
\begin{array}
[c]{lll}%
V(x,c_{1})-V(x,c_{2}) & \leq & J(x;C)+\varepsilon-J(x;\widetilde{C})\\
& = & \mathbb{E}\left[  \int_{0}^{\overline{\tau}}\left(  C_{s}-\widetilde
{C}_{s}\right)  e^{-qs}ds\right]  +\mathbb{E}\left[  \int_{\overline{\tau}%
}^{\tau}C_{s}e^{-qs}ds\right]  +\varepsilon\\
& \leq & \frac{\overline{c}}{q}%
{\textstyle\sum\limits_{j=1}^{\infty}}
\mathbb{E}\left[  I_{\left\{  \overline{\tau}=\tau_{j},\tau>\tau_{j}\right\}
}e^{-q\tau_{j}}\right]  +\varepsilon.
\end{array}
\label{Lipaux1C}%
\end{equation}
Then,%

\[
\mathbb{E}\left[  \left(  e^{-q\overline{\tau}}-e^{-q\tau}\right)  I_{\left\{
\overline{\tau}=\tau_{j}\text{, }\tau>\tau_{j}\right\}  }\right]
\leq\mathbb{E}\left[  e^{-q\tau_{j}}I_{\left\{  \overline{\tau}=\tau
_{j}\text{, }\tau>\tau_{j}\right\}  }\right]  .
\]
Using the definitions given in (\ref{U y At}), we have%

\[%
\begin{array}
[c]{l}%
\left\{  \overline{\tau}=\tau_{j}\text{, }\tau>\tau_{j}\right\} \\%
\begin{array}
[c]{ll}%
= & \left\{  X_{\tau_{j}}^{C}=x+A_{\tau_{j}}^{C}-\mathcal{U}_{j-1}%
\geq0~\text{and }X_{\tau_{j}}^{\widetilde{C}}=x+A_{\tau_{j}}^{\widetilde{C}%
}-\mathcal{U}_{j-1}<0\right\} \\
= & \left\{  x+A_{\tau_{j}}^{C}-\mathcal{U}_{j-1}~\geq U_{j}~>x+A_{\tau_{j}%
}^{\widetilde{C}}-\mathcal{U}_{j-1}\right\} \\
\subseteq & \left\{  x+A_{\tau_{j}}^{\widetilde{C}}+(c_{2}-c_{1})\tau
_{j}-\mathcal{U}_{j-1}~\geq U_{j}~>x+A_{\tau_{j}}^{\widetilde{C}}%
-\mathcal{U}_{j-1}\right\}  .
\end{array}
\end{array}
\]

Note that by the i.i.d.\ assumptions of the compound Poisson process we have
that $\tau_{j}$, $U_{j}$ and $\mathcal{U}_{j-1}$ are mutually independent. Hence,%

\begin{equation}%
\begin{array}
[c]{l}%
\mathbb{E}[\sum_{j=1}^{\infty}e^{-q\tau_{j}}I_{\left\{  \widetilde{\tau}%
=\tau_{j}\text{,}\tau>\tau_{j}\right\}  }]\\%
\begin{array}
[c]{ll}%
\leq & K(c_{2}-c_{1})\beta\sum_{j=1}^{\infty}\left[  \int_{0}^{\infty}%
e^{-qt}\left(  \frac{\beta^{j-1}t^{j-1}}{(j-1)!}\right)  te^{-\beta
t}dt\right] \\
\leq & K\frac{\beta}{q^{2}}(c_{2}-c_{1}),
\end{array}
\end{array}
\label{cota suma esperanzas ruin en tauj}%
\end{equation}
because $F(x+A_{t}^{\widetilde{C}}+(c_{2}-c_{1})t-u)-F(x+A_{t}^{\widetilde{C}%
}-u)\leq(c_{2}-c_{1})t.$ From (\ref{Lipaux1C}) and
(\ref{cota suma esperanzas ruin en tauj}) we get the result with $K_{2}%
=K\beta\overline{c}/q^{3}$.

Let \ us consider now the case $\overline{c}>p$. Take $C\in\Pi_{x,c_{1}%
,\overline{c}}$ as in (\ref{casiOptimac1}) and $\widehat{T}$ as in
(\ref{Definicion TauSombrero}). For
\[
T_{1}:=\min\{t:C_{t}\geq p\},
\]
since $c_{2}\leq p$, we have that $T_{1}\geq\widehat{T}$. Consider the
increasing sequence $(y_{i})$ and the admissible strategies $\widehat{C}%
^{i}\in\Pi_{y_{i},p,\overline{c}}$ introduced in the proof of Lemma
\ref{Lemma Liptchitz x, c menor a p}, and define the dividend payment strategy
$\overline{C}\in\Pi_{x,c_{2},\overline{c}}$ as follows: take rate $c_{2}$ for
$t\leq\widehat{T}$, $C_{t}$ for $\widehat{T}\leq t<T_{1}$ and for $t\geq
T_{1}$ take $\overline{C}_{t}=\widehat{C}_{t-T_{1}}^{i}$ in the case that
$X_{T_{1}}^{C}\in\lbrack y_{i},y_{i+1})$; as before, the strategy
$\overline{C}$ turns out to be Borel measurable and so it is admissible. With
arguments similar to the ones used before, we obtain,%
\[
V(x,c_{1})-V(x,c_{2})\leq\left(  \frac{2}{q}+2\overline{c}~K\frac{\beta}%
{q^{3}}\right)  \left(  c_{2}-c_{1}\right)  \text{. }\blacksquare
\]

\subsection{Proof of Proposition \ref{Proposition Lipschitz cmax mayor a p}}

Proposition \ref{Proposition Lipschitz cmax mayor a p} follows from the
following two lemmas:

\begin{lemma}
\label{Lemma:Lipschx_cmayor a p} Assume that $\overline{c}>p$, then there
exist constants $K_{2\text{ }}>0$ and $K_{3}>0$ such that%
\[
V(x_{2},c)-V(x_{1},c)\leq\left[  K_{2}+\frac{K_{3}}{c-p}\right]  \left(
x_{2}-x_{1}\right)
\]
for all $0\leq x_{1}\leq x_{2}$ and $p<c\leq\overline{c}.$
\end{lemma}

\emph{Proof.} Take $\varepsilon>0$ and $C\in\Pi_{x_{2},c,\overline{c}}$ such that%

\begin{equation}
J(x_{2};C)\geq V(x_{2},c)-\varepsilon. \label{casi optima x2_cmayor a p}%
\end{equation}
Define $\widetilde{C}\in\Pi_{x_{1},c,\overline{c}}$ as $\widetilde{C}%
_{t}=C_{t}$, and let us call $\widetilde{\tau}\leq\tau$ the ruin time of the
process $X_{t}^{\widetilde{C}}$, then $X_{t}^{C}-X_{t}^{\widetilde{C}}%
=x_{2}-x_{1}$ for $t\leq\widetilde{\tau}$. Hence, using
(\ref{casi optima x2_cmayor a p}) and
(\ref{Acotacion_EsperanzaDifx:RuinaSalto}) we have,
\begin{equation}%
\begin{array}
[c]{l}%
V(x_{2},c)-V(x_{1},c)\\%
\begin{array}
[c]{ll}%
= & \mathbb{E}[\int_{\widetilde{\tau}}^{\tau}C_{s}e^{-qs}ds]+\varepsilon\\
\leq & \mathbb{E}[\sum_{j=1}^{\infty}\left(  I_{\left\{  \widetilde{\tau}%
=\tau_{j},\text{ }\tau>\tau_{j}\right\}  }\int_{\tau_{j}}^{\tau}C_{s}%
e^{-qs}ds\right)  ]+\mathbb{E}[\sum_{j=1}^{\infty}\left(  I_{\left\{
\widetilde{\tau}\in(\tau_{j-1,}\tau_{j})\right\}  }\int_{\widetilde{\tau}%
}^{\tau}e^{-qs}C_{s}ds\right)  ]+\varepsilon\\
\leq & \frac{\overline{c}~}{q}\mathbb{E}[\sum_{j=1}^{\infty}e^{-q\tau_{j}%
}I_{\left\{  \widetilde{\tau}=\tau_{j},\tau>\tau_{j}\right\}  }]+\frac
{\overline{c}~}{q}\mathbb{E}[\sum_{j=1}^{\infty}I_{\left\{  \widetilde{\tau
}\in(\tau_{j-1,}\tau_{j})\right\}  }\left(  e^{-q\widetilde{\tau}}-e^{-q\tau
}\right)  ]+\varepsilon.\\
\leq & \overline{c}K\frac{\beta}{q^{2}}(x_{2}-x_{1})+\frac{\overline{c}~}%
{q}\mathbb{E}[\sum_{j=1}^{\infty}I_{\left\{  \widetilde{\tau}\in(\tau
_{j-1,}\tau_{j})\right\}  }\left(  e^{-q\widetilde{\tau}}-e^{-q\tau}\right)
]+\varepsilon.
\end{array}
\end{array}
\label{CotaLipx_cmayorp_aux0}%
\end{equation}
We also get%

\begin{equation}
\mathbb{E}\big[\sum_{j=1}^{\infty}I_{\left\{  \widetilde{\tau}\in(\tau
_{j-1,}\tau_{j})\right\}  }\left(  e^{-q\tau}-e^{-q\widetilde{\tau}}\right)
\big]\leq q\,\mathbb{E}\big[\sum_{j=1}^{\infty}e^{-q\tau_{j-1}}\left(
\tau-\widetilde{\tau}\right)  \big]. \label{CotaLipx_cmayorp_aux2}%
\end{equation}
Assume now that $\widetilde{\tau}\in(\tau_{j-1,}\tau_{j})$ (and so
$\widetilde{\tau}<\tau$). Then
\[
0=X_{\widetilde{\tau}}^{\widetilde{C}}=x_{1}+\int_{0}^{\widetilde{\tau}%
}\left(  p-C_{s}\right)  ds-\sum_{k=1}^{j-1}U_{k}\text{ and }0\leq X_{\tau
^{-}}^{C}\leq x_{2}+\int_{0}^{\tau}\left(  p-C_{s}\right)  ds-\sum_{k=1}%
^{j-1}U_{k}.
\]
Hence, we get%

\[
0\leq X_{\tau^{-}}^{C}-X_{\widetilde{\tau}}^{\widetilde{C}}\leq x_{2}%
-x_{1}+\int_{\widetilde{\tau}}^{\tau}\left(  p-C_{s}\right)  ds\leq
x_{2}-x_{1}+(p-c)(\tau-\widetilde{\tau})
\]
and this implies%

\begin{equation}
\tau-\widetilde{\tau}\leq\frac{x_{2}-x_{1}}{c-p}.
\label{AcotacionDifTau_Lipx_cmayorp}%
\end{equation}
We also have%

\begin{equation}%
\begin{array}
[c]{lll}%
\mathbb{E}[\sum_{j=1}^{\infty}e^{-q\tau_{j-1}}] & = & 1+%
{\displaystyle\int\limits_{0}^{\infty}}
e^{-qs}\beta\sum_{k=1}^{\infty}\left(  \frac{\beta^{k-1}s^{k-1}}%
{(k-1)!}\right)  e^{-\beta s}ds\\
& \leq & 1+\beta/q.
\end{array}
\label{Cuenta con los tauj}%
\end{equation}

So, from (\ref{CotaLipx_cmayorp_aux0}), (\ref{CotaLipx_cmayorp_aux2}),
(\ref{AcotacionDifTau_Lipx_cmayorp}) and (\ref{Cuenta con los tauj}), we get
the result with$~K_{2}=\overline{c}K\beta/q^{2}\ $and $K_{3}=\overline
{c}(1+\beta/q)$. $\blacksquare$

\begin{lemma}
\label{Lema Lipschitz_c_cmayor p} Assume that $\overline{c}>p$, then there
exist constants $K_{2\text{ }}>0$ and $K_{3}>0$ such that%
\[
V(x,c_{1})-V(x,c_{2})\leq\left[  K_{2}+\frac{K_{3}x}{\left(  c_{1}-p\right)
^{2}}\right]  \left(  c_{2}-c_{1}\right)
\]
for all $x\geq0$ and $p<c_{1}\leq c_{2}\leq\overline{c}$.
\end{lemma}

\emph{Proof.} If $x=0,$ $V(x,c)=0$ for all $c>p$. Consider now $x>0$ and
$p<c_{1}<c_{2}\leq\overline{c}$. Take $\varepsilon>0$ and $C\in\Pi
_{x,c_{1},\overline{c}}$ such that $J(x;C)\geq V(x,c_{1})-\varepsilon;$ we
define the admissible strategy
\[
\widehat{T}=\min\{t:C_{t}\geq c_{2}\}.
\]
$\overline{C}\in\Pi_{x,c_{2},\overline{c}}$ as $\overline{C}_{t}%
=c_{2}I_{\left\{  t<\widehat{T}\right\}  }+C_{t}I_{\left\{  t\geq\widehat
{T}\right\}  }$, and the ruin times $\tau$ and $\overline{\tau}$ of the
processes $X_{t}^{C}$ and $X_{t}^{\overline{C}}$ respectively. In this case
both $\tau$ and $\overline{\tau}$ are finite with $\tau\geq\overline{\tau}$.
Note that
\begin{equation}
\tau\leq\frac{x}{c_{1}-p}. \label{Cota_RuinaconC}%
\end{equation}
Let us define as $T_{0}=\min\{t:x+\int_{0}^{t}(p-\overline{C}_{s})ds=0\}$ as
the ruin time of the controlled process $X_{t}^{\overline{C}}$. In the event
of no claims, we have $\overline{\tau}\leq T_{0}$. Since $\overline{c}%
\geq\overline{C}_{s}\geq c_{2}>p$, $T_{0}$ is finite and satisfies%

\begin{equation}
\frac{x}{\overline{c}-p}\leq T_{0}\leq\frac{x}{c_{2}-p}. \label{Cota1T0}%
\end{equation}
So we have
\[
0\leq\int_{0}^{t}\left(  \overline{C}_{s}-C_{s}\right)  ds\leq\left\{
\begin{array}
[c]{ccc}%
(c_{2}-c_{1})t & \text{if} & t\leq\widehat{T}\\
(c_{2}-c_{1})\widehat{T} & \text{if} & t>\widehat{T}%
\end{array}
\right.  ,
\]
and then%
\begin{equation}
X_{\overline{\tau}}^{C}\leq X_{\overline{\tau}^{-}}^{C}\leq X_{\overline{\tau
}^{-}}^{C}-\text{ }X_{\overline{\tau}^{-}}^{\overline{C}}\leq(c_{2}%
-c_{1})\overline{\tau}\leq(c_{2}-c_{1})T_{0}\leq(c_{2}-c_{1})\frac{x}{c_{2}%
-p}. \label{CotasVariasLipschitz}%
\end{equation}
We can write, using (\ref{cota suma esperanzas ruin en tauj}),%
\begin{equation}%
\begin{array}
[c]{l}%
V(x,c_{1})-V(x,c_{2})\\%
\begin{array}
[c]{ll}%
\leq & J(x;C)-J(x;\overline{C})+\varepsilon\\
\leq & \frac{\overline{c}}{q}%
{\textstyle\sum\limits_{j=1}^{\infty}}
\mathbb{E}\left[  I_{\left\{  \overline{\tau}=\tau_{j}\text{, }\tau>\tau
_{j}\right\}  }e^{-q\tau_{j}}\right]  +\frac{\overline{c}}{q}%
{\textstyle\sum\limits_{j=1}^{\infty}}
\mathbb{E}\left[  I_{\left\{  \overline{\tau}\in(\tau_{j-1},\tau_{j})\text{,
}\tau>\overline{\tau}\right\}  }\left(  e^{-q\overline{\tau}}-e^{-q\tau
}\right)  \right]  +\varepsilon\\
\leq & \frac{\overline{c}\beta K}{q^{3}}(c_{2}-c_{1})+\frac{\overline{c}}{q}%
{\textstyle\sum\limits_{j=1}^{\infty}}
\mathbb{E}\left[  I_{\left\{  \overline{\tau}\in(\tau_{j-1},\tau_{j})\text{,
}\tau>\overline{\tau}\right\}  }\left(  e^{-q\overline{\tau}}-e^{-q\tau
}\right)  \right]  +\varepsilon.
\end{array}
\end{array}
\label{Lipaux1C_mayorp}%
\end{equation}
In the case that $\overline{\tau}\in(\tau_{j-1},\tau_{j})$ and $\tau
>\overline{\tau}$, we have that
\[
X_{\overline{\tau}}^{C}+\int_{\overline{\tau}}^{\tau}(p-c_{1})ds\geq
X_{\tau^{-}}^{C}\geq0.
\]
Then we get, from (\ref{CotasVariasLipschitz}),%

\begin{equation}
0\leq\tau-\overline{\tau}\leq\frac{X_{\overline{\tau}}^{C}}{c_{1}-p}\leq
\frac{x}{(c_{1}-p)(c_{2}-p)}(c_{2}-c_{1}). \label{DiferenciasTauTaubarra}%
\end{equation}
Hence, by virtue of (\ref{CotaLipx_cmayorp_aux2}), (\ref{Cuenta con los tauj})
and (\ref{DiferenciasTauTaubarra}),%

\[%
\begin{array}
[c]{l}%
\frac{\overline{c}}{q}%
{\textstyle\sum\limits_{j=1}^{\infty}}
\mathbb{E}\left[  I_{\left\{  \overline{\tau}\in(\tau_{j-1},\tau_{j})\text{,
}\tau>\overline{\tau}\right\}  }\left(  e^{-q\overline{\tau}}-e^{-q\tau
}\right)  \right] \\%
\begin{array}
[c]{ll}%
\leq & \frac{\overline{c}}{q}%
{\textstyle\sum\limits_{j=1}^{\infty}}
\mathbb{E}\left[  I_{\left\{  \overline{\tau}\in(\tau_{j-1},\tau_{j})\text{,
}\tau>\overline{\tau}\right\}  }q\left(  \tau-\overline{\tau}\right)
e^{-q\tau_{j-1}}\right] \\
\leq &
{\textstyle\sum\limits_{j=1}^{\infty}}
\mathbb{E}\left[  I_{\left\{  \overline{\tau}\in(\tau_{j-1},\tau_{j})\text{,
}\tau>\overline{\tau}\right\}  }e^{-q\tau_{j-1}}\right]  \frac{\overline{c}%
x}{(c_{1}-p)(c_{2}-p)}(c_{2}-c_{1})\\
\leq & \frac{\overline{c}x}{(c_{1}-p)(c_{2}-p)}(c_{2}-c_{1})%
{\textstyle\sum\limits_{j=1}^{\infty}}
\mathbb{E}\left[  e^{-q\tau_{j-1}}\right] \\
= & \frac{\overline{c}x}{(c_{1}-p)(c_{2}-p)}\left(  1+\frac{\beta}{q}\right)
(c_{2}-c_{1}).
\end{array}
\end{array}
\]
Therefore, from (\ref{Lipaux1C_mayorp}) the result is established with
$K_{2}=\overline{c}\beta K/q^{3}$ and $K_{3}=\overline{c}\left(
1+\beta/q\right)  $.$~\blacksquare$

\subsection{Proof of Proposition \ref{Continuidad en p en la variable c}}

The proof of Proposition \ref{Continuidad en p en la variable c} is quite
technical. In addition to some technical lemmas below, we will use the
exponential inequality%
\begin{equation}
e^{-\frac{\gamma}{z^{\eta}}\text{ }}\leq\frac{e^{-\frac{1}{\eta}\text{ }}%
}{\left(  \gamma\eta\right)  ^{1/\eta}}z \label{Acotacion Exponencial _eta}%
\end{equation}
for $z>0$, $\gamma>0$ and $\eta>0$, as well as the following elementary remark
about convolutions of independent distribution functions.

\begin{remark}
\normalfont
\label{Lupschitizadad suma de la Ui} The distribution function $F_{j}$ of the
random variable $\mathcal{U}_{j}=U_{1}+...+U_{j\text{ }}$ is Lipschitz with
the same Lipschitz constant as $F$. To see this, consider $\mathcal{U}%
_{2}=U_{1}+U_{2}$. Then%
\[
P(a\leq U_{1}+U_{2}\leq a+h)=\int_{0}^{a+h}\int_{a-u}^{a+h-u}dF(v)dF(u)\leq
Kh\int_{0}^{a+h}dF(u)\leq Kh.
\]
With a recursive argument the proof extends to all $\mathcal{U}_{j}$ for
$j\geq1.$
\end{remark}

Let us call $J_{x}^{c}$ the value function of the strategy in $\Pi
_{x,c,\overline{c}}$ that pays dividends at a constant rate $c$ until ruin. We
first compare $J_{x}^{p}$ with $J_{x}^{c}$ for $c>p$.

\begin{lemma}
\label{Estrategia Constantes Comparacion} If $c>p$, there exists a positive
constant $\overline{K}$, such that,%
\[
-\frac{c-p}{q}\leq J_{x}^{p}-J_{x}^{c}\leq\overline{K}\left[  1+\frac{1}%
{x}+\frac{e^{-\frac{1}{1-\alpha}}}{\left(  xq(1-\alpha)\right)  ^{1/\left(
1-\alpha\right)  }}+\frac{x}{\left(  c-p\right)  ^{1-\alpha}}\right]  (c-p),
\]
for any $0<\alpha<1$ and $x>0$.
\end{lemma}

\emph{Proof.} Let us call $C\in\Pi_{x,p,\overline{c}}$ the constant strategy
$C_{t}=p$ and $\overline{C}\in\Pi_{x,c,\overline{c}}$ the constant strategy
$\overline{C}_{t}=c>p$ for all $t.$ Define again $\tau$ as the ruin time of
the process $X_{t}^{C}$ and $\overline{\tau}$ the one of the process
$X_{t}^{\overline{C}}.$ We have that $\tau$ coincides with the arrival of a
claim and $\overline{\tau}\leq\tau$, so we get the first inequality since%

\[
J_{x}^{c}\leq\int_{0}^{\infty}\left(  c-p\right)  e^{-qs}ds+\int
_{0}^{\overline{\tau}}pe^{-qs}ds\leq\frac{c-p}{q}+J_{x}^{p}.
\]
We can write, using (\ref{cota suma esperanzas ruin en tauj}),%

\begin{equation}%
\begin{array}
[c]{lll}%
J_{x}^{p}-J_{x}^{c} & \leq & \frac{p}{q}\mathbb{E}\left[  e^{-q\overline{\tau
}}-e^{-q\tau}\right]  .\\
& \leq & \frac{p}{q}%
{\textstyle\sum\limits_{j=1}^{\infty}}
\mathbb{E}\left[  I_{\left\{  \overline{\tau}=\tau_{j}\text{, }\tau>\tau
_{j}\right\}  }e^{-q\tau_{j}}\right]  +\frac{p}{q}%
{\textstyle\sum\limits_{j=1}^{\infty}}
\mathbb{E}\left[  I_{\left\{  \overline{\tau}\in(\tau_{j-1},\tau_{j})\text{,
}\tau>\overline{\tau}\right\}  }e^{-q\overline{\tau}}\right] \\
& \leq & \frac{p\beta K}{q^{3}}(c-p)+\frac{p}{q}%
{\textstyle\sum\limits_{j=1}^{\infty}}
\mathbb{E}\left[  I_{\left\{  \overline{\tau}\in(\tau_{j-1},\tau_{j})\text{,
}\tau>\overline{\tau}\right\}  }e^{-q\overline{\tau}}\right]  .
\end{array}
\label{Diferencia de Jotas}%
\end{equation}
Note that if $\overline{\tau}\in(\tau_{j-1},\tau_{j})$, then $\tau
>\overline{\tau}$.

In the event that $\overline{\tau}\in(0,\tau_{1})$ we have $\overline{\tau
}=x/(c-p)$. From (\ref{Acotacion Exponencial _eta}), we get%

\begin{equation}
\mathbb{E}\left[  e^{-q\overline{\tau}}I_{\left\{  \overline{\tau}\in
(0,\tau_{1})\right\}  }\right]  \leq e^{-q\frac{x}{c-p}}\mathbb{E}\left[
I_{\left\{  \overline{\tau}\in(0,\tau_{1})\right\}  }\right]  \leq\frac
{e^{-1}}{qx}(c-p)\mathbb{E}\left[  I_{\left\{  \overline{\tau}\in(0,\tau
_{1})\right\}  }\right]  . \label{Acotacion To}%
\end{equation}

In the event that $\overline{\tau}\in(\tau_{1},\tau_{2})$, we have
$X_{\tau_{1}}^{\overline{C}}=x-(c-p)\tau_{1}-U_{1}>0$. We consider two cases:
$X_{\tau_{1}}^{\overline{C}}\leq x\left(  c-p\right)  ^{\alpha}$ and
$X_{\tau_{1}}^{\overline{C}}>x\left(  c-p\right)  ^{\alpha}.$ In the first
case, using the Lipschitz condition on $F$, we obtain%
\[%
\begin{array}
[c]{l}%
\mathbb{E}\left[  e^{-q\overline{\tau}}I_{\left\{  \overline{\tau}\in(\tau
_{1},\tau_{2})\right\}  }I_{\left\{  0<X_{\tau_{1}}^{\overline{C}}\leq
x\left(  c-p\right)  ^{\alpha}\right\}  }\right] \\%
\begin{array}
[c]{ll}%
\leq & \mathbb{E}\left[  e^{-q\overline{\tau}}I_{\left\{  \overline{\tau}%
\in(\tau_{1},\tau_{2})\right\}  }I_{\left\{  x+(p-c)\tau_{1}-x\left(
c-p\right)  ^{\alpha}\leq U_{1}<x+(p-c)\tau_{1}\right\}  }\right] \\
\leq & \mathbb{E}\left[  e^{-q\tau_{1}}I_{\left\{  \overline{\tau}\in(\tau
_{1},\tau_{2})\right\}  }I_{\left\{  x+(p-c)\tau_{1}-x\left(  c-p\right)
^{\alpha}\leq U_{1}<x+(p-c)\tau_{1}\right\}  }\right] \\
\leq & Kx\left(  c-p\right)  ^{\alpha}\mathbb{E}\left[  e^{-q\tau_{1}%
}I_{\left\{  \overline{\tau}\in(\tau_{1},\tau_{2})\right\}  }\right]  .
\end{array}
\end{array}
\]
In the second case, we have $\left(  \overline{\tau}-\tau_{1}\right)
=X_{\tau_{1}}^{\overline{C}}/(c-p)\geq x/\left(  c-p\right)  ^{1-\alpha}$, and
(\ref{Acotacion Exponencial _eta}) yields%

\[%
\begin{array}
[c]{lll}%
\mathbb{E}\left[  e^{-q\overline{\tau}}I_{\left\{  \overline{\tau}\in(\tau
_{1},\tau_{2})\right\}  }I_{\left\{  X_{\tau_{1}}^{\overline{C}}>x\left(
c-p\right)  ^{\alpha}\right\}  }\right]  & = & \mathbb{E}\left[
e^{-q(\overline{\tau}-\tau_{1})}e^{-q\tau_{1}}I_{\left\{  \overline{\tau}%
\in(\tau_{1},\tau_{2})\right\}  }\right] \\
& \leq & e^{-\frac{qx}{\left(  c-p\right)  ^{1-\alpha}}}\mathbb{E}\left[
e^{-q\tau_{1}}I_{\left\{  \overline{\tau}\in(\tau_{1},\tau_{2})\right\}
}\right]  .
\end{array}
\]
Hence,
\begin{equation}
\mathbb{E}\left[  e^{-q\overline{\tau}}I_{\left\{  \overline{\tau}\in(\tau
_{1},\tau_{2})\right\}  }\right]  \leq\mathbb{E}\left[  e^{-q\tau_{1}}\right]
\left(  Kx\left(  c-p\right)  ^{\alpha}+\frac{e^{-\frac{1}{1-\alpha}}}{\left(
qx(1-\alpha)\right)  ^{1/\left(  1-\alpha\right)  }}(c-p)\right)  .
\label{Acotacion T1}%
\end{equation}
In a similar way, and using Remark \ref{Lupschitizadad suma de la Ui}, we
obtain,
\[
\mathbb{E}\left[  e^{-q\overline{\tau}}I_{\left\{  \overline{\tau}\in
(\tau_{j-1},\tau_{j})\text{, }\tau>\overline{\tau}\right\}  }\right]
\leq\mathbb{E}\left[  e^{-q\tau_{j}}\right]  \left(  Kx\left(  c-p\right)
^{\alpha}+\frac{e^{-\frac{1}{1-\alpha}}}{\left(  xq(1-\alpha)\right)
^{1/\left(  1-\alpha\right)  }}(c-p)\right)
\]
for any $j\geq3$ and so from (\ref{Cuenta con los tauj}),
(\ref{Diferencia de Jotas}) and (\ref{Acotacion To}) we get the second
inequality.\hfill$~\blacksquare$\newline

In the next lemma, we give an alternative version of the Lipschitz condition
for $x>0$ and $c>p$. Here, for $x_{2}>x_{1}\geq\delta>0$, the growth of the
Lipschitz bound as $c\rightarrow p^{+}$ , goes to infinity but slower than the
bound obtained in Lemma \ref{Lemma:Lipschx_cmayor a p}.

\begin{lemma}
\label{Proposition Lipschitz cmax mayor a p_Alternativa} For any $0<\alpha<1$
there exists a positive constant $\widetilde{K}$ such that%
\[
V(x_{2},c)-V(x_{1},c)\leq\widetilde{K}\left[  1+\frac{1}{x_{1}}+\frac
{e^{-\frac{1}{1-\alpha}}}{\left(  qx_{1}(1-\alpha)\right)  ^{1/\left(
1-\alpha\right)  }}+\frac{x_{1}}{\left(  c-p\right)  ^{1-\alpha}}\right]
(x_{2}-x_{1}),
\]
where $p<c\leq\overline{c}$ and $0<x_{1}<x_{2}.$
\end{lemma}

\emph{Proof.} Take $\varepsilon>0$ and $C\in\Pi_{x_{2},c,\overline{c}}$ such that%

\begin{equation}
J(x_{2};C)\geq V(x_{2},c)-\varepsilon\label{casi optima x2_cmayor a p_Alt}%
\end{equation}
and call $\tau$ the ruin time of the process $X_{t}^{C}$. Define
$\widetilde{C}\in\Pi_{x_{1},c,\overline{c}}$ as $\widetilde{C}_{t}=C_{t}$ and
call $\widetilde{\tau}$ the ruin time of the process $X_{t}^{\widetilde{C}}$;
it holds that $\widetilde{\tau}\leq\tau$ and $X_{t}^{C}-X_{t}^{\widetilde{C}%
}=x_{2}-x_{1}$ for $t\leq\widetilde{\tau}$. In the event that $\widetilde
{\tau}\in(\tau_{j-1,}\tau_{j})$ (and so $\widetilde{\tau}<\tau$),
$X_{\widetilde{\tau}}^{\widetilde{C}}=0$ and so $X_{\widetilde{\tau}}%
^{C}=X_{\widetilde{\tau}}^{\widetilde{C}}+(x_{2}-x_{1})=x_{2}-x_{1}$. Hence,
since $C_{s}\geq C_{\widetilde{\tau}}$ for $s\geq\widetilde{\tau},$%

\begin{equation}
\tau-\widetilde{\tau}\leq\frac{1}{C_{\widetilde{\tau}}-p}\int_{\widetilde
{\tau}}^{\tau}\left(  C_{s}-p\right)  ds\leq\frac{x_{2}-x_{1}}{C_{\widetilde
{\tau}}-p}. \label{AcotacionDifTau_Lipx_cmayorp_Alt}%
\end{equation}
From (\ref{CotaLipx_cmayorp_aux0}) and (\ref{AcotacionDifTau_Lipx_cmayorp_Alt}%
), we get%

\begin{equation}%
\begin{array}
[c]{l}%
V(x_{2},c)-V(x_{1},c)\\%
\begin{array}
[c]{ll}%
\leq & \overline{c}K\frac{\beta}{q^{2}}(x_{2}-x_{1})+\frac{\overline{c}~}%
{q}\mathbb{E}[\sum_{j=1}^{\infty}I_{\left\{  \widetilde{\tau}\in(\tau
_{j-1,}\tau_{j})\right\}  }e^{-q\widetilde{\tau}}\left(  1-e^{-q(\tau
-\widetilde{\tau})}\right)  ]+\varepsilon\\
\leq & \overline{c}K\frac{\beta}{q^{2}}(x_{2}-x_{1})+\overline{c}%
\mathbb{E}[\sum_{j=1}^{\infty}I_{\left\{  \widetilde{\tau}\in(\tau_{j-1,}%
\tau_{j})\right\}  }e^{-q\widetilde{\tau}}\frac{1}{C_{\widetilde{\tau}}%
-p}]\left(  x_{2}-x_{1}\right)  +\varepsilon
\end{array}
\end{array}
\label{desigualdad V No Lipschitz}%
\end{equation}
since $1-e^{-ay}\leq ay$.

In the event that $\widetilde{\tau}\in(0,\tau_{1}),$%

\[
(C_{\widetilde{\tau}}-p)\widetilde{\tau}\geq\int_{0}^{\widetilde{\tau}}\left(
C_{s}-p\right)  ds=x_{1}\text{, }%
\]
so $\widetilde{\tau}\geq x_{1}/(C_{\widetilde{\tau}}-p)$. By
(\ref{Acotacion Exponencial _eta}), we get%

\begin{equation}
\mathbb{E}\left[  \frac{e^{-q\overline{\tau}}}{C_{\widetilde{\tau}}%
-p}I_{\left\{  \overline{\tau}\in(0,\tau_{1})\right\}  }\right]  \leq
\frac{e^{-1\text{ }}}{qx_{1}}. \label{Acotacion caso 1 con tau}%
\end{equation}

In the event that $\widetilde{\tau}\in(\tau_{1},\tau_{2})$, we consider two
cases: $X_{\tau_{1}}^{\widetilde{C}}>x_{1}\left(  C_{\widetilde{\tau}%
}-p\right)  ^{\alpha}$ and $0<X_{\tau_{1}}^{\widetilde{C}}\leq x_{1}\left(
C_{\widetilde{\tau}}-p\right)  ^{\alpha}$. Analogously to the proof of Lemma
\ref{Estrategia Constantes Comparacion}, we use the Lipschitz condition on $F$
in the first case and (\ref{Acotacion Exponencial _eta}) in the second case to obtain%

\[%
\begin{array}
[c]{ccc}%
\mathbb{E}\left[  I_{\left\{  \widetilde{\tau}\in(\tau_{1,}\tau_{2}\right\}
}\frac{e^{-q\widetilde{\tau}}}{C_{\widetilde{\tau}}-p}\right]  & = &
\mathbb{E}\left[  \frac{e^{-q\widetilde{\tau}}}{C_{\widetilde{\tau}}%
-p}I_{\left\{  \widetilde{\tau}\in(\tau_{1},\tau_{2})\right\}  }%
I_{0<X_{\tau_{1}}\leq x_{1}\left(  C_{\widetilde{\tau}}-p\right)  ^{\alpha}%
}\right] \\
&  & +\mathbb{E}\left[  \frac{e^{-q\widetilde{\tau}}}{C_{\widetilde{\tau}}%
-p}I_{\left\{  \widetilde{\tau}\in(\tau_{1},\tau_{2})\right\}  }I_{X_{\tau
_{1}}>x_{1}\left(  C_{\widetilde{\tau}}-p\right)  ^{\alpha}}\right] \\
& \leq & \left(  \frac{Kx_{1}}{\left(  c-p\right)  ^{1-\alpha}}+\frac
{e^{-\frac{1}{1-\alpha}}}{\left(  qx_{1}(1-\alpha)\right)  ^{1/\left(
1-\alpha\right)  }}\right)  \mathbb{E}\left[  e^{-q\tau_{1}}\right]  .
\end{array}
\]
In a similar way, and using Remark \ref{Lupschitizadad suma de la Ui}, we
obtain
\[
\mathbb{E}\left[  I_{\left\{  \widetilde{\tau}\in(\tau_{j-1,}\tau
_{j})\right\}  }e^{-q\widetilde{\tau}}\frac{1}{C_{\widetilde{\tau}}-p}\right]
\leq\left(  \frac{Kx_{1}}{\left(  c-p\right)  ^{1-\alpha}}+\frac{e^{-\frac
{1}{1-\alpha}}}{\left(  qx_{1}(1-\alpha)\right)  ^{1/\left(  1-\alpha\right)
}}\right)  \mathbb{E}\left[  e^{-q\tau_{j-1}}\right]
\]
for any $j\geq3$ and so from (\ref{Cuenta con los tauj}),
(\ref{Acotacion caso 1 con tau}) and (\ref{desigualdad V No Lipschitz}), we
get the result. \hfill$\blacksquare$\newline

\emph{Proof of Proposition \ref{Continuidad en p en la variable c}.} Consider
$x>0$, we need to prove that $\lim_{c\rightarrow p^{+}}V(x,c)=V(x,p)$. Let us
call, as before, $J_{y}^{c}$ the value function of the strategy in
$\Pi_{y,c,\overline{c}}$ that pays dividends at a constant rate $c$ until
ruin. Then, by Remark \ref{Remark Pagar p}, $V(0,p)=J_{0}^{p}$. Also, we get
$0\leq J_{y}^{p}-J_{0}^{p}=J_{y}^{p}-V(0,p)$, and from Proposition
\ref{Proposition Global Lipschitz zone} there exists a $K_{1}>0$ such that
$V(y,p)-V(0,p)\leq K_{1}y.$ Hence,%

\[
V(y,p)-J_{y}^{p}\leq V(y,p)-V(0,p)+V(0,p)-J_{y}^{p}\leq K_{1}y.
\]
So, given $\varepsilon>0$ small enough and taking $\delta\leq\varepsilon
/K_{1}$, we have
\begin{equation}
V(y,p)-J_{y}^{p}\leq\varepsilon\label{Cotap_ychico}%
\end{equation}
for all initial surplus levels $0\leq y\leq\delta$. We assume $\delta
<\min\left\{  1/4,x\right\}  $, so $\delta^{3/2}<\delta/2$. Consider $C\in
\Pi_{x,p,\overline{c}}$ such that $J(x;C)\geq V(x,p)-\varepsilon$ and define
$T_{1}:=\min\left\{  t\geq0:X_{t}^{C}\leq\delta\right\}  $ and $T_{2}$ such that%

\[
\int_{T_{2}}^{\infty}e^{-qs}\overline{c}ds=\frac{\overline{c}}{q}e^{-qT_{2}%
}\leq\varepsilon\text{.}%
\]
Take $c\in$ $(p,\overline{c})$ such that
\begin{equation}
c-p\leq\min\{\delta^{3/2}/T_{2},\left(  \varepsilon/T_{2}\right)
^{5},\varepsilon,\delta^{3/2}\}. \label{Cota c-p}%
\end{equation}
Let us define $\widehat{T}:=\min\{t:C_{t}\geq c\}$. Since $V(\cdot,c)$ is
non-decreasing and continuous, we can find (as in Lemma
\ref{Lemma Liptchitz x, c menor a p}) an increasing sequence $(y_{i})\ $with
$y_{1}=0$ such that if $y\in\lbrack y_{i},y_{i+1})$, then $0\leq
V(y,c)-V(y_{i},c)\leq\varepsilon/2$. Consider admissible strategies
$\widehat{C}^{i}\in\Pi_{y_{i},c,\overline{c}}$ such that $V(y_{i}%
,c)-J(y_{i},\widehat{C}^{i})\leq\varepsilon/2$.

Let us now define the admissible strategy $\overline{C}\in\Pi_{x,c,\overline
{c}}$ as follows: $\overline{C}_{t}=c$ for $t<\widehat{T}$ ; in the event that
$T_{1}\leq\widehat{T}$ (and so $X_{\widehat{T}}^{C}\leq\delta)$, the strategy
for $t\geq\widehat{T}$ consists of paying dividends at constant rate $c$ until
ruin; and in the event that $T_{1}>\widehat{T}$ (and so $X_{\widehat{T}}%
^{C}>\delta),$ we define $\overline{C}_{t}=\widehat{C}_{t-T_{1}}^{i}$ for
$t\geq\widehat{T}$ in the case that $X_{T_{1}}^{C}\in\lbrack y_{i},y_{i+1})$.
Note that with this definition the strategy $\overline{C}$ turns out to be
admissible and $C_{s}-\overline{C}_{s}\leq0$ for $s\leq\widehat{T}$.

Let us call $\tau$ and $\overline{\tau}$ the ruin times of the processes
$X_{t}^{C}$ and $X_{t}^{\overline{C}}$, respectively. In order to prove the
result, we consider different cases depending on the value of $\widehat{T}$:%

\begin{equation}%
\begin{array}
[c]{l}%
V(x,p)-V(x,c)\\%
\begin{array}
[c]{ll}%
\leq & J(x;C)-J(x;\overline{C})+\varepsilon~\\
= & \mathbb{E}\left[  I_{\left\{  \widehat{T}\geq\overline{\tau}\right\}
}(J(x;C)-J(x;\overline{C}))\right]  +\mathbb{E}\left[  I_{\left\{  \widehat
{T}<\overline{\tau},\widehat{T}>T_{2}\right\}  }(J(x;C)-J(x;\overline
{C}))\right] \\
& +\mathbb{E}\left[  I_{\left\{  \widehat{T}<\overline{\tau},\widehat{T}\leq
T_{2}\wedge T_{1}\right\}  }(J(x;C)-J(x;\overline{C}))\right]  +\varepsilon\\
& +\mathbb{E}\left[  I_{\left\{  \widehat{T}<\overline{\tau},\widehat{T}%
\in\left[  T_{1},T_{2}\right]  \right\}  }(I_{\left\{  T_{1}\neq\tau
_{j}\text{, }1\leq j\right\}  }+%
{\textstyle\sum\limits_{j=1}^{\infty}}
I_{\left\{  T_{1}=\tau_{j}\right\}  })\left(  J(x;C)-J(x;\overline{C})\right)
\right]  .
\end{array}
\end{array}
\label{Expresion Basica}%
\end{equation}

In the event $\widehat{T}\geq\overline{\tau}$, using $\tau\geq\overline{\tau}$
and Lemma \ref{Estrategia Constantes Comparacion}, we can show that
\begin{equation}%
\begin{array}
[c]{l}%
\mathbb{E}\left[  I_{\left\{  \widehat{T}\geq\overline{\tau}\right\}
}(J(x;C)-J(x;\overline{C}))\right] \\
\leq\overline{K}\left[  1+\frac{1}{x}+\frac{e^{-\frac{1}{1-\alpha}}}{\left(
xq(1-\alpha)\right)  ^{1/\left(  1-\alpha\right)  }}+\frac{x}{\left(
c-p\right)  ^{1-\alpha}}+\frac{1}{q}\right]  (c-p).
\end{array}
\label{Expresion Basica 0}%
\end{equation}

In the event that $\widehat{T}<\overline{\tau}\ $and $\widehat{T}>T_{2}$, by
the definition of $T_{2}$,
\begin{equation}
\mathbb{E}\left[  I_{\left\{  \widehat{T}<\overline{\tau},\widehat{T}%
>T_{2}\right\}  }(J(x;C)-J(x;\overline{C}))\right]  \leq\mathbb{E}\left[
\int_{T_{2}}^{\infty}e^{-qs}\overline{c}ds\right]  \leq\varepsilon.
\label{Expresion Basica 1}%
\end{equation}

In the event that $\widehat{T}<\overline{\tau},\widehat{T}\leq T_{2}\wedge
T_{1}$, it holds that $X_{\widehat{T}}^{C}\geq\delta$ and
\[
0\leq X_{\widehat{T}}^{C}-X_{\widehat{T}}^{\overline{C}}\leq(c-p)\widehat
{T}<(c-p)T_{2}\leq\min\left\{  \varepsilon,\delta^{3/2}\right\}
<\delta/2\text{.}%
\]
Therefore, since $V(\cdot,c)$ is non-decreasing and $X_{\widehat{T}%
}^{\overline{C}}\in\lbrack\delta/2,x)$, we obtain from Lemma
\ref{Proposition Lipschitz cmax mayor a p_Alternativa}%

\begin{equation}%
\begin{array}
[c]{l}%
\mathbb{E}\left[  I_{\left\{  \widehat{T}<\overline{\tau},\widehat{T}\leq
T_{2}\wedge T_{1}\right\}  }(J(x;C)-J(x;\overline{C}))\right] \\%
\begin{array}
[c]{ll}%
\leq & \widetilde{K}\left(  1+\frac{2}{\delta}+\frac{e^{-\frac{1}{1-\alpha}}%
}{\left(  q(1-\alpha)\delta/2\right)  ^{1/\left(  1-\alpha\right)  }}\right)
\delta^{3/2}+\widetilde{K}x(c-p)^{\alpha}T_{2}+\varepsilon.
\end{array}
\end{array}
\label{Expresion Basica 2}%
\end{equation}

In the event that $\widehat{T}<\overline{\tau}$ and $\widehat{T}\geq T_{1}$,
the strategy is $\overline{C}_{t}=c$ for all $t$. If $T_{1}$ does not coincide
with the arrival of a claim, then $X_{T_{1}}^{C}=\delta$ (and so $X_{T_{1}%
}^{\overline{C}}\geq\delta/2)$. Then we can write, using (\ref{Cotap_ychico}),
Proposition \ref{Proposition Global Lipschitz zone} and Lemma
\ref{Estrategia Constantes Comparacion},%

\begin{equation}%
\begin{array}
[c]{l}%
\mathbb{E}\left[  I_{\left\{  \widehat{T}<\overline{\tau},\widehat{T}%
\in\left[  T_{1},T_{2}\right]  \right\}  }I_{\left\{  T_{1}\neq\tau_{j}\text{,
}1\leq j\right\}  }(J(x;C)-J(x;\overline{C}))\right] \\%
\begin{array}
[c]{ll}%
\leq & \mathbb{E}\left[  I_{\left\{  \widehat{T}<\overline{\tau},\widehat
{T}\in\left[  T_{1},T_{2}\right]  \right\}  }I_{\left\{  T_{1}\neq\tau
_{j}\text{, }1\leq j\right\}  }e^{-qT_{1}}(V\left(  \delta,p\right)
-J_{\delta/2}^{c})\right] \\
\leq & \left(  V\left(  \delta,p\right)  -J_{\delta}^{p}\right)  +\left(
J_{\delta}^{p}-J_{\delta/2}^{p}\right)  +\left(  J_{\delta/2}^{p}-J_{\delta
/2}^{c}\right) \\
\leq & \varepsilon+K_{1}\frac{\varepsilon}{2K_{1}}+\overline{K}(1+\frac
{2}{\delta}+\frac{e^{-\frac{1}{1-\alpha}}}{\left(  \delta/2q(1-\alpha)\right)
^{1/\left(  1-\alpha\right)  }})\delta^{3/2}+\overline{K}\delta/2(c-p)^{\alpha
}~.
\end{array}
\end{array}
\label{Expresion Basica 3}%
\end{equation}

Finally, in the event that $\widehat{T}<\overline{\tau},$ $\widehat{T}\geq
T_{1}$ and $T_{1}$ coincides with the $j$-th claim arrival, then $X_{T_{1}%
}^{C}=X_{\tau_{j}}^{C}\in(0,$ $\delta)$ and $X_{\tau_{j}^{-}}^{C}\geq$
$\delta$. Hence,%
\[
0<X_{T_{1}}^{C}=X_{\tau_{j}}^{C}=X_{\tau_{j}^{-}}^{C}-U_{j}<\delta.
\]
Therefore, $X_{\tau_{j}^{-}}^{C}>U_{j}~>X_{\tau_{j}^{-}}^{C}-\delta\geq0$.
Since $F(X_{\tau_{j}^{-}}^{C})-F(X_{\tau_{j}^{-}}^{C}-\delta)\leq K\delta$
and, by the compound Poisson assumptions, we obtain%
\[
\mathbb{E}\left[  I_{\left\{  \widehat{T}<\overline{\tau},\widehat{T}%
\in\left[  T_{1},T_{2}\right]  \right\}  }I_{\left\{  T_{1}=\tau_{j}\right\}
}e^{-q\tau_{j}}\right]  \leq K\delta\mathbb{E}\left[  e^{-q\tau_{j}}\right]
.
\]
So, by (\ref{Cuenta con los tauj}) and Proposition
\ref{Proposition Lipschitz cmax mayor a p},%
\begin{equation}%
\begin{array}
[c]{l}%
{\textstyle\sum\limits_{j=1}^{\infty}}
\mathbb{E}\left[  I_{\left\{  \widehat{T}<\overline{\tau},\widehat{T}%
\in\left[  T_{1},T_{2}\right]  \right\}  }I_{\left\{  T_{1}=\tau_{j}\right\}
}(J(x;C)-J(x;\overline{C}))\right] \\%
\begin{array}
[c]{ll}%
\leq & K\delta V(\delta,c)\sum_{j=1}^{\infty}\mathbb{E}\left[  e^{-q\tau_{j}%
}\right]  \ \\
\leq & \frac{K\overline{c}\beta}{q^{2}}\delta.
\end{array}
\end{array}
\label{Expresion Basica 4}%
\end{equation}
Using (\ref{Cota c-p})--(\ref{Expresion Basica 4}) with $\alpha=1/5$, and so
$1/\left(  1-\alpha\right)  =5/4<3/2,$ we get the result.\hfill$\blacksquare$

\bigskip

\end{document}